\documentclass[aps,prl,twocolumn,superscriptaddress]{revtex4-2}

\usepackage{amsthm}
\usepackage{amsmath,bm}
\usepackage{amssymb}
\usepackage{amsfonts}
\usepackage{graphicx}
\usepackage{txfonts}
\usepackage{xcolor}
\usepackage{float}
\usepackage{braket}

\usepackage[colorlinks=true,linkcolor=blue,citecolor=blue,urlcolor=blue]{hyperref}

\begin{document}

\title{Detecting Bell correlations in multipartite non-Gaussian spin states}

\author{Jiajie Guo}
\affiliation{State Key Laboratory for Mesoscopic Physics, School of Physics, Frontiers Science Center for Nano-optoelectronics, $\&$ Collaborative
Innovation Center of Quantum Matter, Peking University, Beijing 100871, China}

\author{Jordi Tura}
\affiliation{Instituut-Lorentz, Universiteit Leiden, P.O. Box 9506, 2300 RA Leiden, The Netherlands}

\author{Qiongyi He}
\email{qiongyihe@pku.edu.cn}
\affiliation{State Key Laboratory for Mesoscopic Physics, School of Physics, Frontiers Science Center for Nano-optoelectronics, $\&$ Collaborative
Innovation Center of Quantum Matter, Peking University, Beijing 100871, China}
\affiliation{Collaborative Innovation Center of Extreme Optics, Shanxi University, Taiyuan, Shanxi 030006, China}
\affiliation{Peking University Yangtze Delta Institute of Optoelectronics, Nantong 226010, Jiangsu, China}

\author{Matteo Fadel}
\email{fadelm@phys.ethz.ch}
\affiliation{Department of Physics, ETH Z\"{urich}, 8093 Z\"{urich}, Switzerland}

\begin{abstract}
We expand the toolbox for studying Bell correlations in multipartite systems by introducing permutationally invariant Bell inequalities (PIBIs) involving few-body correlators. First, we present around twenty families of PIBIs with up to three- or four-body correlators, that are valid for arbitrary number of particles. Compared to known inequalities, these show higher noise robustenss, or the capability to detect Bell correlations in highly non-Gaussian spin states. We then focus on finding PIBIs that are of practical experimental implementation, in the sense that the associated operators require collective spin measurements along only a few directions. To this end, we formulate this search problem as a semidefinite program that embeds the constraints required to look for PIBIs of the desired form.
\end{abstract}

\maketitle

Some correlations arising from quantum physics cannot be explained within the paradigm of local realism, and are thus called nonlocal \cite{BrunnerRMP2014}. These are detected via the violation of a so-called Bell inequality \cite{Bell1964}, tested in practice through a Bell experiment \cite{HensenNature2015}. Besides their fundamental interest, nonlocal correlations are the resource enabling device-independent (DI) quantum information processing tasks, such as quantum key distribution \cite{AcinDIQKD}, randomness amplification \cite{ColbeckRenner2012} or self-testing \cite{SupicQuantum2020}.
Although much research has focused on few-partite scenarios, mostly bipartite \cite{PitowskyPRA2001, Faacets}, nonlocal correlations also appear naturally in the multipartite regime \cite{MerminPRL1990, TothPRA2006} and, in particular, in physically relevant many-body systems \cite{SciencePaper, AnnPhys, Kitzinger21}. With mild additional assumptions, multipartite nonlocality can be revealed in experimentally-practical ways, and take the name of Bell correlations \cite{SchmiedScience2016, EngelsenPRL2017}.

Detection of Bell correlations is of great interest, as they are related to quantum critical points \cite{PigaPRL2019}, metrology \cite{FroewisPRAR2019}, open quantum systems \cite{MarconiPRAR2022}, and bosonic systems at finite temperature \cite{FadelQuantum2018, NiezgodaPRA2019}, and provide an avenue to quantify DI entanglement and Bell correlation depth \cite{AloyPRL2019, TuraPRA2019, BaccariPRA2019}. However, the available inequalities are scarce, because a complete characterization is an intractable task \cite{BabaiCompCompl1991}. An approach that finds a good compromise between expressivity and complexity is to focus on Bell inequalities with particular symmetries and low-order correlators \cite{SciencePaper, AnnPhys, TIpaper, TuraPRX2017}. In turn, this reduces the experimental requirements to reveal Bell correlations from them. A paradigmatic example is the use of two-body, permutationally invariant Bell inequalities (PIBIs) to detect a class of Gaussian states known as spin-squeezed states \cite{SchmiedScience2016, EngelsenPRL2017}.

Despite all this progress, so far only PIBIs with up to two-body correlators are known, which poses a fundamental limit on their applicability. It is thus of great interest to find PIBIs involving higher-order moments of physical observables, such that Bell correlations can be detected in larger classes of states and with higher noise tolerance. In particular, these tools would enable the study of Bell correlations in non-Gaussian states \cite{WalschaersPRXQ2021}, which cannot be characterized by only second moments.

In this work we present around twenty new PIBIs involving three- and four-body correlators, and illustrate that compared to known PIBIs they provide an advantage in terms of noise robustness and sensitivity to non-Gaussian states. Moreover, we provide a general framework to derive new PIBIs with high-order correlators under the constraint of being experimentally practical, in the sense that they can be tested by performing collective spin measurements along only a few directions. This is based on a semidefinite program (SDP) that allows us to find Bell correlation witnesses of a desired ansatz.

\vspace{2mm}
\textbf{Preliminaries.--}
We consider the multipartite Bell experiment in which N observers, labeled as $i=1,2,\cdots,N$, perform one of the two measurements $M_0^{(i)}, M_1^{(i)}$, on their part of the system, and obtain one of the two possible outcomes $\pm 1$. Correlations among parties are characterized by the correlators $\langle M_{j_1}^{(i_1)} \cdots M_{j_k}^{(i_k)}\rangle $. However, to reduce the complexity of characterising multipartite correlations, we restrict ourselves to permutationally invariant (PI) observables, namely
\begin{align}\label{eq:corr}
\mathcal{S}_{j_1\cdots j_k} &= \sum_{\stackrel{1\leq i_1<\cdots<i_k\leq N}{\text{+ perm.}}} \langle M_{j_1}^{(i_1)} \cdots M_{j_k}^{(i_k)}\rangle \;.
\end{align}
Here, $k=1,...,K$ indicates the order of the PI correlator, and $j_l=0,1$ is the measurement setting for $l=1,\ldots, k$.

By considering $N$ parties and at most $K$th-order PI correlators, the set of classical correlations form a polytope $\mathbb{P}_{N,K}^S$.
The vertices of this polytope are identified with the correlations originating from deterministic local hidden variable models (LHVM), for which it holds $\langle M_{j_1}^{(i_1)}\cdots M_{j_k}^{(i_k)} \rangle =\langle M_{j_1}^{(i_1)}\rangle \cdots \langle M_{j_k}^{(i_k)} \rangle$ and $\langle M_j^{(i)} \rangle =\pm 1$ for all $i, j$.
From these vertices, it is possible to derive a dual description of $\mathbb{P}_{N,K}^S$ in terms of the linear inequalities (i.e. PIBIs) defining its facets.
Measurement statistics lying outside $\mathbb{P}_{N,K}^S$ indicate the presence of Bell correlations. For details about this framework we refer the reader to Refs.~\cite{SciencePaper,AnnPhys}.

In a scenario with $N<20$ and $K\lesssim 3$ it is relatively simple to list all vertices of $\mathbb{P}_{N,K}^S$ and to obtain from them the full list of facet inequalities. For larger $N$ and $K$, however, this approach is unfeasible. Therefore, to find PIBIs that allow the detection of Bell correlations in many-body systems, we might rely on the following method.
For a given $K$, we characterize $\mathbb{P}_{N,K}^S$ for a few small values of $N$. Then, we look for patterns in the inequalities that appear, and use them to conjecture families of PIBIs valid for arbitrary $N$. Finally, we prove the conjectured families by demonstrating that they cannot be violated by LHVM.

Until now, only a couple of PIBIs valid for arbitrary $N$ are known, and only for $K=2$ \cite{SciencePaper,WagnerPRL2017}. Of particular relevance is the inequality
\begin{equation}\label{JTBell6}
    I_2 \equiv -2\mathcal{S}_0+\frac{1}{2}\mathcal{S}_{00}-\mathcal{S}_{01}+\frac{1}{2}\mathcal{S}_{11} + 2N \geq 0 \;,
\end{equation}
which enabled the experimental detection of Bell correlations in spin-squeezed BECs \cite{SchmiedScience2016} and cold atomic ensembles \cite{EngelsenPRL2017}.

\vspace{2mm}
\textbf{Third-order Bell inequalities.--}
We start with considering the case $K=3$. Remarkably, by computing the polytope $\mathbb{P}_{N,3}^S$ for all $N<12$, we were able to identify around twenty families of Bell inequalities (See Supplementary Sec. V). As we will see, an interesting family among these is
\begin{align} \label{MFBell}
& I_3 \equiv -12(N-1)\mathcal{S}_0 -12(N-1)\mathcal{S}_1  + 3(N-2)\mathcal{S}_{00} + 6 N \mathcal{S}_{01} + \notag\\
&\;  3(N-2)\mathcal{S}_{11} -2 \mathcal{S}_{000} -3 \mathcal{S}_{001} + \mathcal{S}_{111} + 12 N (N-1) \geq 0 
\end{align}
which we have proven to be valid for all $N$ in Supplementary Sec. I.

As all the inequalities we consider here involve two measurement settings and two outcomes, Jordan's lemma guarantees that their maximum quantum violation can be achieved by qubit measurements \cite{Toner2006, JordanBook}. To this end, we identify $M_j^{(i)} = \vec{u}_j \cdot \vec{\sigma}^{(i)} \;(j=0,1)$ for the $i$th party, where $\vec{\sigma}$ is the vector of Pauli matrices. Even if not necessarily optimal nor required by the inequality, we assume that the same pair of observables is chosen by all parties, i.e. $M_j^{(i)}=M_{j}$. Since local rotations are irrelevant here, we can simplify further our discussion by choosing $M_0=\sigma_z$ and $M_1(\theta)= \sin(\theta) \,\sigma_x + \cos(\theta) \, \sigma_z$, 
where $\theta\in[0,\pi]$. With this definition, the correlators Eq.~\eqref{eq:corr} can be written as operators
\begin{align}
\hat{\mathcal{S}}_{j_1\cdots j_k} (\theta)&= \sum_{\stackrel{1\leq i_1<\cdots<i_k\leq N}{\text{+ perm.}}}  M_{j_1}^{(i_1)}(\theta) \otimes \cdots \otimes M_{j_k}^{(i_k)}(\theta) \;,
\end{align}
so that the Bell inequalities (\ref{JTBell6}) and (\ref{MFBell}) can now be understood as Bell operators $\hat{I}_2(\theta)$ and $\hat{I}_3(\theta)$, respectively.

Given a Bell operator, we search for the optimal measurements and states that maximizes the quantum violation relative to the classical bound, i.e. $Q_V^N/\beta_C^N$, where the classical bound $\beta_C^N$ is the constant term appearing in the inequalities. Bell nonlocality is detected for a state $|\psi\rangle$ if the Bell operator yields a negative expectation value $\langle \psi|\hat{I}_K|\psi\rangle<0$. As the classical bound is a constant, the maximum quantum violation $Q_V^N$ can thus be identified with the minimum eigenvalue of Bell operator
\begin{align}\label{eq:mineigv}
    Q_V^N = \min_\theta \min_{|\psi\rangle} \langle \psi|\hat{I}_K|\psi\rangle =\min_\theta \lambda_{min}(\hat{I}_K) \;,
\end{align}
and the associated eigenvector is the state that maximally violates $I_K$. 

\begin{figure}[t]
	\begin{center}
		\includegraphics[width=85mm]{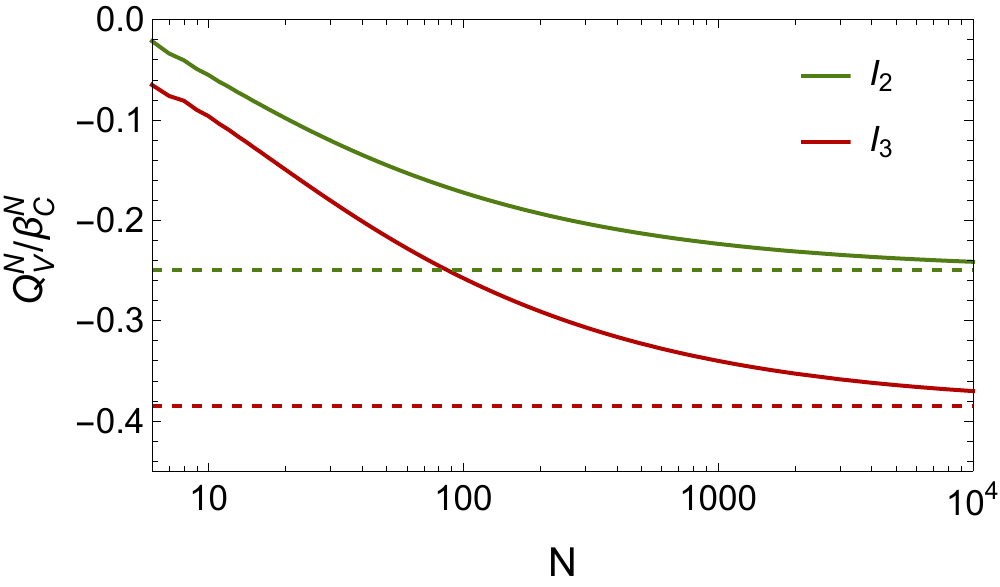}
	\end{center}
	\caption{Maximum relative quantum violation $Q_V^N/\beta_C^N$ for the 3rd-order Bell inequality $I_3$ and 2nd-order Bell inequality $I_2$, as a function of the number of parties $N$. The two horizontal dashed lines indicate the asymptotic violation for $N\rightarrow\infty$, which for $I_2$ is $-1/4$, and for $I_3$ is $-2\sqrt{3}/9\approx -0.3849$. 
	}
	\label{Fig1}
\end{figure}

In the $N$-qubit Hillbert space, the dimension of Bell operator scales exponentially with $N$, making challenging to solve the eigenvalue problem Eq.~\eqref{eq:mineigv} for large $N$. Fortunately, since the correlators $\mathcal{S}$ have permutation symmetry, it is possible to introduce a symmetry-adapted basis in which to express the Bell operator, such that it block-diagonalizes due to Schur-Weyl duality \cite{FultonHarris1999, MoroderNJP2012}. We can then focus the search of nonlocality onto the fully symmetric block $\hat{I}_K^\text{Sym}$ of size $(N+1)\times(N+1)$, and the maximum quantum violation of $\hat{I}_K$ can then be obtained from the lowest eigenvalue of $\hat{I}_K^\text{Sym}$ \cite{AnnPhys}. In Fig.\ref{Fig1} we show $Q_V^N/\beta_C$ for $I_2$ and $I_3$, as a function of $N$. It is evident the significantly better scaling for the higher-order Bell inequalities $I_3$ compared to $I_2$. In the limit $N\rightarrow\infty$, it is possible to show through a variational calculation \cite{AnnPhys} that the relative violation of $I_3$ tends to $-2\sqrt{3}/9\approx -0.3849$, which is larger than the value $-1/4$ obtained for $I_2$. This indicates a higher noise robustness, as well as the possibility to detect Bell correlations in a larger class of states, as we will see with an example in the next paragraph. A similar analysis for the quantum violation of some other third-order Bell inequalities $I_3$ and a fourth-order Bell inequality $I_4$ are given in Supplementary Sec. V, VI.

\vspace{2mm}
\textbf{Bell correlations in spin-squeezed states.--}
We now show that $I_3$ allows us to detect Bell correlations in many-body spin states of experimental relevance. To illustrate this, let us consider the states that can be prepared through the paradigmatic one-axis twisting (OAT) Hamiltonian $H=\hbar\chi \hat{S}_z^2$ \cite{KitagawaPRA1993}. The evolution of an initial coherent spin state along the $x$-axis for a time $t$ can be parametrized through the (adimensional) interaction strength $\mu=2\chi t$, and reads $|\Phi(\mu)\rangle= 2^{-N/2} \sum_{k=0}^N \sqrt{\binom{N}{k}} e^{-i (N/2-k)^2 \mu / 2}|k\rangle$, 
where $|k\rangle$ represents the $N$-particle Dicke state with $k$ excitations.

To investigate Bell correlations in the OAT states $|\Phi(\mu)\rangle$, we compute $I_2$ and $I_3$ as a function of $\mu$. For this, we need to minimize the associated Bell operators over the measurement directions $M_j$ at every $\mu$, which in the case of a given state have to be both parametrized over the full sphere as $M_j = \vec{u}_j \cdot \vec{\sigma}$, with $\vec{u}_j=(\cos(\phi_j) \sin(\theta_j),\sin(\phi_j) \sin(\theta_j),\cos(\theta_j))$. This yields Bell operators $\hat{I}_K(\phi_0,\theta_0,\phi_1,\theta_1)$ that are now functions of four angles.

\begin{figure}[t]
	\begin{center}
		\includegraphics[width=85mm]{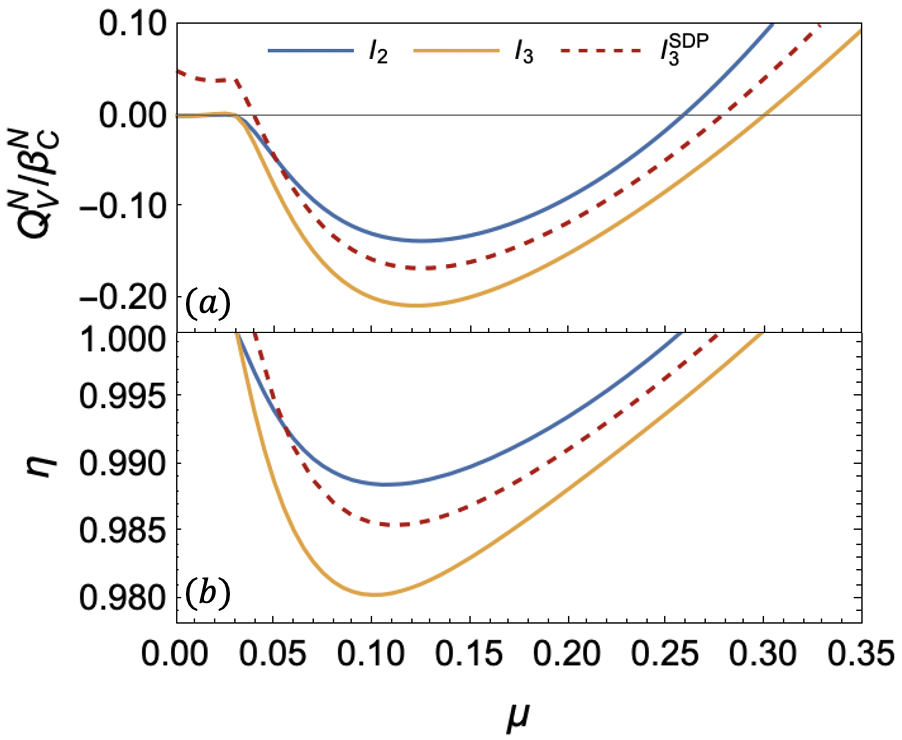}
	\end{center}
	\caption{a): Relative quantum violation of the PIBIs $I_2$ (blue) and $I_3$ (orange) for $N=50$ spin OAT states $|\psi(\mu)\rangle$ as a function of $\mu$. An advantage over $I_2$ can also be found for $I_3^{\text{SDP}}$ (red dashed), which requires to measure only one third moment of the collective spin. b): For mixed states $\rho(\eta,\mu)=\eta \ket{\Phi(\mu)}\bra{\Phi(\mu)}+(1-\eta)\mathbf{I}/(N+1)$, the minimum purity $\eta$ required to violate each PIBI.}
	\label{Fig2new}
\end{figure}

Concretely, we express the Bell operators in the fully-symmetric subspace in terms of collective spin operators, make use of the analytic results for their expectation values for spin OAT states (see Supplementary Sec. II, III), and then minimize the lowest eigenvalue of the operator over the measurement directions.
The violations we obtain are shown in Fig.~\ref{Fig2new} for $N=50$, where we can observe that $I_3$ outperforms $I_2$ by reaching larger relative violation $Q_V^N/\beta_C^N$ as well as detecting Bell correlations over a wider squeezing range, and thus for a larger class of states. To investigate the noise robustness, we consider OAT states mixed with white noise as $\rho(\eta,\mu)=\eta \ket{\Phi(\mu)}\bra{\Phi(\mu)}+(1-\eta)\mathbf{I}/(N+1)$. In Fig.~\ref{Fig2new} we plot the minimum $\eta$ for observing a PIBI violation, and show that high-order inequalities detect Bell correlations with higher noise tolerance. 

\vspace{2mm}
\textbf{Finding practical high-order Bell inequalities.--}
We have shown, taking $I_3$ as an example, that high-order Bell inequalities can outperform their low-order counterparts and allow us to robustly detect Bell correlations in states that are routinely investigated experimentally. However, we note that the associated Bell operator often requires to measure high-order moments of the collective spin along several directions, because we have e.g.  (see Supplementary Sec. II)
\begin{equation}
     \mathcal{S}_{001} = \frac{8}{3} \langle \hat{S}_{\vec{n}}\hat{S}_{\vec{n}}\hat{S}_{\vec{m}}+\hat{S}_{\vec{n}}\hat{S}_{\vec{m}}\hat{S}_{\vec{n}}+\hat{S}_{\vec{m}}\hat{S}_{\vec{n}}\hat{S}_{\vec{n}}\rangle + ... \;. 
\end{equation}
This can pose practical challenges, as estimating many high-order moments requires to collect large measurement statistics. For this reason, we would like to find inequalities with coefficients satisfying additional constraints, such that the associated operator involves e.g. only \emph{one} third-order moment. To see the form of such constraint, note that for some measurement direction $\vec{a}=\alpha \vec{m} + \beta \vec{n}$ we have (see Supplementary Sec. IV)
\begin{equation} \label{eq:onedirectionS3}
    \langle \hat{S}^3_{\vec{a}} \rangle = \frac{\beta^3}{8}\mathcal{S}_{000} + \frac{\alpha^3}{8}\mathcal{S}_{111} + \frac{3\alpha \beta^2}{8}\mathcal{S}_{001} + \frac{3\alpha^2\beta}{8} \mathcal{S}_{011} + f(\mathcal{S}_0, \mathcal{S}_1) \;,
\end{equation}
where $f(\mathcal{S}_0, \mathcal{S}_1)$ is a linear function of one-body correlators only, and $\alpha,\beta\in\mathbb{R}$, $|\alpha^2|+|\beta^2|=1$.
Therefore, only inequalities whose coefficients for the third-order correlators are following the pattern of Eq.~\eqref{eq:onedirectionS3} will result in a Bell operator involving only one third-order moment of the collective spin.

Note here that imposing such constraints as a further projection of the local polytope is not trivial, because of the high nonlinearity in $\alpha$ and $\beta$ of the coefficients multiplying the correlators. One obtains a different polytope projection for each pair $(\alpha, \beta)$, which dramatically increases the optimization complexity, as for each projection new families of Bell inequalities need to be found. Moreover, such approach does not guarantee to find tight inequalities, as these might correspond to tilts of facets.

We thus propose a way to circumvent these difficulties, by developing a method to find practical Bell operators that is based on a hierarchy of semidefinite programs (SDPs) that search for inequalities of a particular form. For the sake of brevity and clarity, in the following we present our method in brief, and show a concrete example for $K=3$. For a more detailed and general formulation we refer the reader to Supplementary Sec. VII.

To illustrate our idea, let us start recalling that for deterministic LHVM the PI correlators Eq.~\eqref{eq:corr} can be written as polynomials in four non-negative integers $(a,b,c,d)$ such that $a+b+c+d=N$ (see Supplementary Sec. I). This allows us to parametrize the vertices of $\mathbb{P}_{N,K}^S$ with integer partitions of $N$, but it also implies that considering $(a,b,c,d)$ to be non-negative reals gives us a outer approximation of $\mathbb{P}_{N,K}^S$ in terms of a semialgebraic set. In the space of PI correlators, the latter is specified by: (i) the set of polynomial equalities $f_i(\vec{\cal S})=0$ expressing constraints among correlators, e.g. ${\cal S}_{000} = {\cal S}_0^3-(3(\mathcal{S}_0^2-\mathcal{S}_{00})-2){\cal S}_0$, and (ii) the set of four polynomial inequalities $g_i(\vec{\cal S})\geq 0$ expressing $(a,b,c,d)\geq 0$ through correlators, e.g. $a\geq 0$ implies $g_1(\vec{\cal S})=(\mathcal{S}_0^2-\mathcal{S}_{00})+\mathcal{S}_0+\mathcal{S}_1+(\mathcal{S}_0\mathcal{S}_1-\mathcal{S}_{01}) \geq 0$.

Having at hand an outer approximation of $\mathbb{P}_{N,K}^S$ in terms of a semialgebraic set, we can apply known techniques that use SDP hierarchies to test membership to the convex hull of such set \cite{ParriloBook2013, Gouveia2010, Gouveia2012}. The first step consists in defining a basis vector, e.g. $\vec{b}=(1,\mathcal{S}_0,\cdots,\mathcal{S}_{111})^T$, from which the moment matrix $\tilde{\Gamma}=\displaystyle\oplus_{i=1}^4 g_i\vec{b}\cdot\vec{b}^T$ is constructed. Then, to check whether the point $\vec{\mathcal{S}}^*$ is outside the convex hull approximating $\mathbb{P}_{N,K}^S$ we can linearize $\tilde{\Gamma}$ and write the SDP
\begin{equation}
    \begin{array}{cccc}
        \max_{\tilde{\Gamma}} &\lambda&\\
        \mathrm{s.t.}&\tilde{\Gamma} &\succeq &0\\
        &\tilde{\Gamma}_{00} &= &1\\
        &\tilde{\Gamma}_{0i} &=& \lambda (\vec{\mathcal{S}}^*)_i\\
        &\tilde{\Gamma}_{ij} &=& p(\tilde{\Gamma}) \;,
    \end{array}
    \label{eq:SdPmain}
\end{equation}
where $p(\tilde{\Gamma})$ is the function expressing the constraints between the entries of $\tilde{\Gamma}$. If SDP~\eqref{eq:SdPmain} returns $\lambda<1$ we must conclude that $\vec{\mathcal{S}}^*_3$ lies outside the convex hull outer approximating $\mathbb{P}_{N,K}^S$, and therefore that this point cannot be described by a LHVM. In this case, the SDP dual to \eqref{eq:SdPmain} gives us a certificate of nonlocality by providing a PIBI that is violated by $\vec{\mathcal{S}}^*_3$.

To ensure that the dual to \eqref{eq:SdPmain} provides us with PIBIs that are experimentally practical, we now modify SDP~\eqref{eq:SdPmain} by adding the additional constraints required to obtain Bell operators of the desired form. Note however that this is nontrivial, as such constraints are highly nonlinear (cf. Eq.~\eqref{eq:onedirectionS3}). For example, asking for a  Bell operator involving only one third moment of the collective spin requires to impose the constraints
\begin{align} \label{eq:addConstr}
        (\tilde{\Gamma}_{01}, \cdots, \tilde{\Gamma}_{05})&= \lambda(\mathcal{S}_0^*,\mathcal{S}_1^*,\cdots,\mathcal{S}_{11}^*) \\
        y &= \lambda(\beta^3\mathcal{S}_{000}^*+3\alpha\beta^2\mathcal{S}_{001}^*+3\alpha^2\beta \mathcal{S}_{011}^*+\alpha^3\mathcal{S}_{111}^*)  \notag\;,
\end{align}
where $y=(\beta^3 \tilde{\Gamma}_{06} +3\alpha\beta^2\tilde{\Gamma}_{07}+3\alpha^2\beta \tilde{\Gamma}_{08}+\alpha^3\tilde{\Gamma}_{09})$. Therefore, we run SDP~\eqref{eq:SdPmain} with Eq.~\eqref{eq:addConstr} as an optimization problem over $(\alpha,\beta)$, in order to find $\min_{\alpha,\beta}(\lambda)$.

\vspace{2mm}
\textbf{Example with OAT states.--}
Let us go back to the problem of detecting Bell correlations in OAT spin states. We aim to find a third-order PIBI with coefficients such that the resulting Bell operator requires the measurement of only one third moment of the collective spin.

First, we specify the target state $|\Phi(\mu)\rangle$ where to detect Bell correlations, e.g. by choosing $N=50$ and $\mu=0.2$. Then, we find a pair of measurement axes $\vec{u}_{1,2}$ such that the resulting list of correlators $\vec{\mathcal{S}}^*_3=(\mathcal{S}_0^*,\mathcal{S}_1^*,\mathcal{S}_{00}^*,\cdots,\mathcal{S}_{111}^*)$ shows Bell correlations. This step can be implemented by using SDP~\eqref{eq:SdPmain} to run the optimization problem $\min_{\vec{u}_{1},\vec{u}_{2}}(\lambda)$. At this point, note that the dual to SDP~\eqref{eq:SdPmain} will provide us a PIBI that is violated by $\vec{\mathcal{S}}^*_3$, but that is in general of difficult experimental implementation.
For this reason, we modify SDP~\eqref{eq:SdPmain} to include the constraints Eq.~\eqref{eq:addConstr} and run the optimization problem $\min_{\alpha,\beta}(\lambda)$. The dual SDP gives now a PIBI in the form 
\begin{align}
    I_3^{\text{SDP}} = c_0 + c_1\mathcal{S}_0+c_2\mathcal{S}_1+\cdots+c_{6}\left(\beta^3 \mathcal{S}_{000} + ... \right) \geq 0 \;,
\end{align}
where $\vec{c}=(c_0,\cdots,c_5,c_6)$ are the variables dual to $(\tilde{\Gamma}_{00},\cdots,\tilde{\Gamma}_{05},y)$ respectively, and whose associated Bell operator involves only one third-order moment of the collective spin.
For our target state, we obtain $\alpha/\beta=41/59$ and $\vec{c}=(1,-0.0055,-0.0141,0.0046,0.0099,0.0051,-56.1412)$. The resulting $I_3^{\text{SDP}}$ has a worse noise tolerance than $I_3$, but it still outperforms $I_2$, see Fig.~\ref{Fig2new}. This could be improved by searching for a different inequality for each $\mu$, or by allowing the measurement of two third-order moments.

\begin{figure}
    \centering
    \includegraphics[width=\columnwidth]{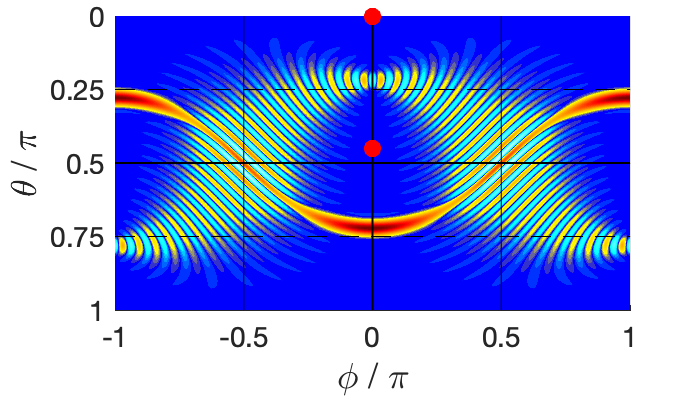}
    \caption{For $N=50$, Wigner function of the eigenstate corresponding to the minimum eigenvalue of Bell operator Eq.~\eqref{eq:I4op}. Red dots indicate the optimal measurement direction $\hat{S}_{\vec{n}}, \hat{S}_{\vec{m}}$, for violating $I_4$.}
    \label{fig:3}
\end{figure}

\vspace{2mm}
\textbf{Fourth-order Bell inequalities.--}
Following the ideas presented so far, we can now guide our search for experimentally practical high-order PIBIs. For $K=4$ we find the inequality
\begin{align}\label{eq:I4}
&I_4 \equiv 24(N-1)\mathcal{S}_{00}+48(N-1)\mathcal{S}_{01}+24(N-3)\mathcal{S}_{11}+ \\ 
&+\mathcal{S}_{0000}+4\mathcal{S}_{0001}+6\mathcal{S}_{0011}+4\mathcal{S}_{0111}+\mathcal{S}_{1111}+48N(N-1)\geq 0 \notag
\end{align}
which involves two- and four-body correlators. The associate Bell operator reads
\begin{align} \label{eq:I4op}
    \hat{I}_4 &= 64 \hat{S}^4_{\vec{a}}  -192  \hat{S}^2_{\vec{m}} +32 \, g(\vec{a},\vec{m}) \, \hat{S}^2_{\vec{a}} + h(\vec{a},\vec{m})
\end{align}
where $g$ and $h$ are scalar functions of $\vec{a}\cdot\vec{m}$ (see Supplementary Sec. VI). Remarkably, note that Eq.~\eqref{eq:I4op} involves measurements of the collective spin operator along two directions only. Diagonalising $\hat{I}_4$ according to Eq.~\eqref{eq:mineigv} we further conclude that: (i) its maximum relative quantum violation increases with $N$, (ii) the states maximally violating inequality \eqref{eq:I4} are highly non-Gaussian, and resembling a superposition of OAT states, see Fig.~\ref{fig:3}. The latter states do not violate $I_2$, as it is also expected from the fact that they have zero polarisation (see Supplementary Sec. VIII).

\vspace{2mm}
\textbf{Conclusions.--} We addressed the problem of finding multipartite PIBIs involving correlators of order higher than second, and that are of practical experimental implementation in the sense that the associated operators require collective measurements along only a few directions. We propose about twenty new PIBIs valid for arbitrary $N$ that involve up to three-body correlators, and one that involves two- and four-body correlators. From a systematic analysis, we conclude that in general these inequalities indeed outperform the currently known PIBIs, since they show higher noise tolerance and the ability to detect Bell correlations in highly non-Gaussian states. In general, for a PIBI to be experimentally practical, we note that the coefficients of the correlators must satisfy some (nonlinear) constraints. We find that these can be imposed a priori, and formulate a SDP that looks for PIBIs resulting in Bell operators of the desired form (e.g. involving only one third-moment). Our results can pave the way to studying Bell correlations in non-Gaussian spin states, and to use generalized spin-squeezing parameters as Bell correlation witnesses.

\vspace{2mm}
\textit{Acknowledgments.--} This work is supported by the National Natural Science Foundation of China (Grants No. 12125402 and No. 11975026).  JG acknowledges financial support from the China Scholarship Council (Grant No. 202106010192). QH acknowledges the Beijing Natural Science Foundation (Z190005) and the Key R$\&$D Program of Guangdong Province (Grant No. 2018B030329001). MF was supported by The Branco Weiss Fellowship -- Society in Science, administered by the ETH Z\"{u}rich.

\bibliographystyle{apsrev4-1} 
\bibliography{HighOrderPIBI_v3.bbl}

\clearpage
\newpage

\begin{widetext}

\section{Supplemental material for ``Detecting Bell correlations in multipartite non-Gaussian spin states''}

\section{I.\quad  Expressions of $\mathcal{S}$ for local deterministic strategies}\label{si:sectionSLDS}

For our scenario, in which parties have two measurement settings $\{M_0,M_1\}$ and two possible outcomes $\{\pm 1\}$, only four local deterministic strategies (LDS) are possible, corresponding to the combinations $\langle M_{0(1)} \rangle = \pm 1$. Since PIBIs are invariant under permutation of the parties, it is not important which party follows which strategy, but only the number of parties following each of the strategies. For this reason, we define the four non-negative integer variables
\begin{align}\label{eq:abcd}
    & a = \# \{i\in \{1,\cdots,N\} \,|\, \langle M_0^{(i)} \rangle=1, \langle M_1^{(i)} \rangle=1\}, \nonumber \\  
    & b = \# \{i\in \{1,\cdots,N\} \,|\, \langle M_0^{(i)} \rangle=1, \langle M_1^{(i)} \rangle=-1\}, \nonumber \\ 
    & c = \# \{i\in \{1,\cdots,N\} \,|\, \langle M_0^{(i)} \rangle=-1,\langle M_1^{(i)} \rangle=1\}, \nonumber \\ 
    & d = \# \{i\in \{1,\cdots,N\} \,|\, \langle M_0^{(i)} \rangle=-1,\langle M_1^{(i)} \rangle=-1\}, 
\end{align}
counting the number of parties using each one of the four strategies, such that $a+b+c+d=N$.

Using the fact that for LDS correlators factorise as 
\begin{align}
    \langle M^{(i)}_{j_1} \cdots M^{(j)}_{j_k} \rangle_{\text{LDS}} = \langle M^{(i)}_{j_1}\rangle \cdot \dots \cdot \langle M^{(j)}_{j_k} \rangle \;,
\end{align}
we can express PI correlators $\mathcal{S}$ as polynomials in $\{a,b,c,d\}$. For example, one-body correlators read
\begin{align}
    \mathcal{S}_k=\sum_{i=1}^N \langle M^{(i)}_k \rangle \;,
\end{align}
such that we have the polynomials
\begin{align}
    \mathcal{S}_0 &= 
    a +b -c -d \;, \\
    \mathcal{S}_1 &=a-b+c-d \;.
\end{align}
For the two-body correlators we can write
\begin{align}\label{eq:decM2nd}
    \mathcal{S}_{kl} &=\sum_{i,j=1(i\neq j)}^N \langle M^{(i)}_k M^{(j)}_l \rangle = \sum_{i,j=1}^N \langle M^{(i)}_k M^{(j)}_l \rangle - \sum_{i=1}^N \langle M^{(i)}_k M^{(i)}_l \rangle \nonumber \\
    &= \sum_{i=1}^N \langle M^{(i)}_k \rangle \sum_{j=1}^N \langle M^{(j)}_l \rangle - \sum_{i=1}^N \langle M^{(i)}_k \rangle \langle M^{(i)}_l \rangle \;,
\end{align}
which gives the polynomials
\begin{align}
    \mathcal{S}_{00} 
    &= \mathcal{S}^2_0-N, \\
    \mathcal{S}_{11} &= \mathcal{S}^2_1-N, \\
    \mathcal{S}_{01} 
    &= \mathcal{S}_0 \mathcal{S}_1-(a-b-c+d)
\end{align}
Performing the same decomposition as Eq.~\eqref{eq:decM2nd} for three- and four-body correlators we get
\begin{align}
    \mathcal{S}_{000} 
    &= \mathcal{S}^3_0+2\mathcal{S}_0-3N\mathcal{S}_0 \label{eq:S000} \\
    \mathcal{S}_{111} &= \mathcal{S}^3_1+2\mathcal{S}_1-3N\mathcal{S}_1 \\
    \mathcal{S}_{001} 
    &= \mathcal{S}_0\mathcal{S}_0\mathcal{S}_1 +2\mathcal{S}_1 -N\mathcal{S}_1-2(a-b-c+d)\mathcal{S}_0 \\
    \mathcal{S}_{011} 
    &= \mathcal{S}_0\mathcal{S}_1\mathcal{S}_1 +2\mathcal{S}_0-N\mathcal{S}_0-2(a-b-c+d)\mathcal{S}_1 . \label{eq:S011} \\
    \mathcal{S}_{0000} 
&= \mathcal{S}_0\mathcal{S}_0\mathcal{S}_0\mathcal{S}_0-6N+3N^2-6N\mathcal{S}_0\mathcal{S}_0+8\mathcal{S}_0\mathcal{S}_0 \\
\mathcal{S}_{1111} &= \mathcal{S}_1\mathcal{S}_1\mathcal{S}_1\mathcal{S}_1-6N+3N^2-6N\mathcal{S}_1\mathcal{S}_1+8\mathcal{S}_1\mathcal{S}_1 \\
\mathcal{S}_{0001} &=\mathcal{S}_0\mathcal{S}_0\mathcal{S}_0\mathcal{S}_1-6(a-b-c+d)+3N(a-b-c+d)-3N\mathcal{S}_0\mathcal{S}_1 -3(a-b-c+d)\mathcal{S}_0\mathcal{S}_0+8\mathcal{S}_0\mathcal{S}_1 \\
\mathcal{S}_{0111} &= \mathcal{S}_0\mathcal{S}_1\mathcal{S}_1\mathcal{S}_1-6(a-b-c+d)+3N(a-b-c+d)-3N\mathcal{S}_0\mathcal{S}_1 -3(a-b-c+d)\mathcal{S}_1\mathcal{S}_1+8\mathcal{S}_0\mathcal{S}_1 \\
\mathcal{S}_{0011} &= \mathcal{S}_0\mathcal{S}_0\mathcal{S}_1\mathcal{S}_1-6N+N^2+2(a-b-c+d)^2-N\mathcal{S}_1\mathcal{S}_1-N\mathcal{S}_0\mathcal{S}_0 -4(a-b-c+d)\mathcal{S}_0\mathcal{S}_1+4\mathcal{S}_1\mathcal{S}_1+4\mathcal{S}_0\mathcal{S}_0 \;.
\end{align}

\subsection{Proof of the classical bound for $I_3$}\label{si:section1}

Here we prove that $I_3$ is a valid Bell inequality for arbitrary $N$, by showing that it cannot be violated by LHVM. 
The results presented in the previous section allow us to rewrite the Bell inequality $I_3$ of the main text as a polynomial in the $\{a,b,c,d\}$ variables, namely
\begin{align}
    I_3(a,b,c,d) = 8(a-d)(a-d-1)(3c+2d+a-2)+48b(c+d) \;.
\end{align}
If $I_3(a,b,c,d)\geq 0$ for all integer partitions of $N$ into $\{a,b,c,d\}$, then we must conclude that $I_3$ cannot be violated by LHVM, and thus that it is a valid Bell inequality.

Before starting with the proof, let us note that since $b,c,d\geq 0$, we have
\begin{align}
    I_3(a,b,c,d) \geq 8(a-d)(a-d-1)(3c+2d+a-2) \;.
\end{align}
Let us then begin by proving that 
\begin{align}
    (a-d)(a-d-1)\geq 0, \;\; \forall(a,d)\in \mathbb{Z}_{\geq 0 }^2 \;.
\end{align}
This can be seen as an equation in $e\equiv a-d$ which reads
\begin{align}
    e(e-1)=(e-1/2)^2-1/4,\;\; e\in\mathbb{Z} \;,
\end{align}
which means that $e(e-1)\geq 0$ for all integer $e$: It is strictly positive if $e>1$ from the sum of squares decomposition and zero if $e=0$ or $e=1$. It does not matter that the minimum over the reals is $e=-1/2$ for which the expression evaluates to $-1/4<0$.

Hence, it remains to prove that $3c+2d+a-2\geq0$ on nonnegative integers. Note, however, that we do not need to consider the cases where $a-d=0$ or $a-d-1=0$, since these are prefactors that will make the whole expression zero.

Since $c\geq0$, clearly $3c+2d+a-2 \geq 2d+a-2$. This last expression can be negative on the following pairs $(a,d)$:
\begin{itemize}
    \item $(a,d)=(0,0)$. This is discarded by the premultiplication by $(a-d)$, which evaluates the whole thing to zero.
    
    \item $(a,d)=(1,0)$. This is again discarded by the premultiplication by $(a-d-1)$, which also eventually evaluates to zero.
    
    \item If $a>1$ or $d>0$, then $2d+a-2\geq0$, so there are no more remaining cases.
\end{itemize}
All the possibilities are exhausted, thus completing the proof that 
\begin{align}
    I_3(a,b,c,s)\geq 0 \quad \forall(a,b,c,d) \in \mathbb{Z}_{\geq0}^4 \;.
\end{align}
This result implies that $I_3$ is a valid Bell inequality, as it shows that it cannot be violated by LHVM. 
Following the same approach as the one presented here, one can prove the validity of any other PIBIs.

\clearpage
\newpage
\section{II.\quad  Writing $\mathcal{S}$ correlators in terms of collective spin expectation values}\label{si:section2}

Our goal is here to show how PI correlators $\mathcal{S}$ can be measured from collective spin measurements.
First, we parametrise our measurements $M_k$ as spin projection measurements, namely
\begin{align}
    M^{(i)}_k=\vec{u}_k \cdot \vec{\sigma}^{(i)} \;
\end{align}
with $\vec{u}_k\in\mathbb{R}^3$ and $|\vec{u_k}|^2=1$. Here, $k=0,1$ labels one of the two measurement settings considered, and $\vec{\sigma}=\{\sigma_x,\sigma_y,\sigma_z\}$ is the vector of Pauli matrices. In the following, we will use the relabeling $\vec{u}_0=\vec{n}$ and $\vec{u}_1=\vec{m}$.

We consider the case where the same pair of observables are chosen by all parties, i.e.  $M_k^{(i)}=M_k$ independent of $i$. This allows us to write 
\begin{align}
    \mathcal{S}_k = \sum_{i=1}^N \langle M^{(i)}_k \rangle = 2 \hat{S}_{\vec{u}_k} \;,
\end{align}
where we have introduced the $N$-particle collective spin operator along direction $\vec{u}_k$ as
\begin{equation}
    \hat{S}_{\vec{u}_k} = \dfrac{1}{2} \sum_{i=1}^N \vec{u}_k \cdot \vec{\sigma}^{(i)} \;.
\end{equation}
The one-body correlators thus read
\begin{align}
    \mathcal{S}_0 &=
    2 \langle \hat{S}_{\vec{n}} \rangle ,\\
    \mathcal{S}_1 &= 2 \langle \hat{S}_{\vec{m}} \rangle \;.
\end{align}

Following the same idea, we can express correlators of any order as collective spin measurements. For example, for the two-body correlators we can write
\begin{align}\label{eq:2ndS}
    \mathcal{S}_{kl}=\sum_{i,j=1(i\neq j)}^N \langle M_k^{(i)} M_l^{(j)} \rangle = \sum_{i,j=1}^N \langle M_k^{(i)} M_l^{(j)} \rangle - \sum_{i=1}^N \langle M_k^{(i)} M_l^{(i)} \rangle \;.
\end{align}
which gives
\begin{align}
    \mathcal{S}_{00} &= 
    4\langle \hat{S}_{\vec{n}}^2 \rangle - N ,\\
    \mathcal{S}_{11} &= 4\langle \hat{S}_{\vec{m}}^2 \rangle - N ,\\
    \mathcal{S}_{01} &= 2 \langle \hat{S}_{\vec{n}} \hat{S}_{\vec{m}}+\hat{S}_{\vec{m}} \hat{S}_{\vec{n}} \rangle - (\vec{m}\cdot \vec{n}) N \;.
\end{align}

The three-body correlators can be expressed as
\begin{align}
    \mathcal{S}_{klm} &= \sum^N_{i,j,y=1(i\neq j \neq y)} \langle M_k^{(i)} M_l^{(j)} M_m^{(y)} \rangle \nonumber \\
    &= \sum^N_{i,j,y=1} \langle M_k^{(i)} M_l^{(j)} M_m^{(y)} \rangle + 2\sum^N_{i=1} \langle M_k^{(i)} M_l^{(i)} M_m^{(i)} \rangle \nonumber \\
    & \;- \langle \sum^N_{i=1} M^{(i)}_k M^{(i)}_l \sum^N_{y=1} M^{(y)}_m \rangle - \langle \sum^N_{j=1} M^{(j)}_l M^{(j)}_m \sum^N_{i=1} M^{(i)}_k \rangle - \langle \sum^N_{y=1} M^{(y)}_m M^{(y)}_k \sum^N_{j=1} M^{(j)}_l \rangle \;,
\end{align}
which gives
\begin{align}
    \mathcal{S}_{000} 
    &= 8\langle \hat{S}^3_{\vec{n}} \rangle + 4 \langle \hat{S}_{\vec{n}} \rangle - 6N \langle \hat{S}_{\vec{n}} \rangle  , \\
    \mathcal{S}_{111} &= 8\langle \hat{S}^3_{\vec{m}} \rangle + 4 \langle \hat{S}_{\vec{m}} \rangle - 6N \langle \hat{S}_{\vec{m}} \rangle ,\\
    \mathcal{S}_{001} &= \frac{1}{3} \left[8\langle \hat{S}_{\vec{n}}\hat{S}_{\vec{n}}\hat{S}_{\vec{m}}+\hat{S}_{\vec{n}}\hat{S}_{\vec{m}}\hat{S}_{\vec{n}}+\hat{S}_{\vec{m}}\hat{S}_{\vec{n}}\hat{S}_{\vec{n}}\rangle+(4-6N)\langle \hat{S}_{\vec{m}} \rangle + (8-12N)(\vec{n}\cdot \vec{m}) \langle \hat{S}_{\vec{n}} \rangle \right] ,\\
    \mathcal{S}_{110} &= \frac{1}{3} \left[8\langle \hat{S}_{\vec{m}}\hat{S}_{\vec{m}}\hat{S}_{\vec{n}}+\hat{S}_{\vec{m}}\hat{S}_{\vec{n}}\hat{S}_{\vec{m}}+\hat{S}_{\vec{n}}\hat{S}_{\vec{m}}\hat{S}_{\vec{m}}\rangle+(4-6N)\langle \hat{S}_{\vec{n}} \rangle + (8-12N)(\vec{n}\cdot \vec{m}) \langle \hat{S}_{\vec{m}} \rangle \right] \;.
\end{align}

Finally, the four-body correlators can be expressed as
\begin{align}
    \mathcal{S}_{klmn} &= \sum_{i,j,y,h=1(i\neq j\neq y\neq h)} \langle M^{(i)}_k M^{(j)}_l M^{(y)}_m M^{(h)}_n \rangle \nonumber \\
    &= \langle \sum^N_{i,j,y,h=1} M^{(i)}_k M^{(j)}_l M^{(y)}_m M^{(h)}_n \rangle -6 \langle \sum^N_{i=1} M^{(i)}_k M^{(i)}_l M^{(i)}_m M^{(i)}_n \rangle \nonumber \\
    &\;+ \langle \sum^N_{i=1} M^{(i)}_k M^{(i)}_l \sum^N_{y=1} M^{(y)}_m M^{(y)}_n \rangle + \langle \sum^N_{i=1} M^{(i)}_k M^{(i)}_m \sum^N_{j=1} M^{(j)}_l M^{(j)}_n \rangle + \langle \sum^N_{i=1} M^{(i)}_k M^{(i)}_n \sum^N_{j=1} M^{(j)}_l M^{(j)}_m \rangle \nonumber \\
    & \; -\langle \sum^N_{i=1} M^{(i)}_k M^{(i)}_l \sum^N_{y=1} M^{(y)}_m \sum^N_{h=1} M^{(h)}_n \rangle -\langle \sum^N_{i=1} M^{(i)}_k M^{(i)}_m \sum^N_{j=1} M^{(j)}_l \sum^N_{h=1} M^{(h)}_n  \rangle -\langle \sum^N_{i=1} M^{(i)}_k M^{(i)}_n \sum^N_{j=1} M^{(j)}_l \sum^N_{y=1} M^{(y)}_m  \rangle \nonumber \\
    & \; -\langle \sum^N_{j=1} M^{(j)}_l M^{(j)}_m \sum^N_{i=1} M^{(i)}_k \sum^N_{h=1} M^{(h)}_n  \rangle -\langle \sum^N_{j=1} M^{(j)}_l M^{(j)}_n \sum^N_{i=1} M^{(i)}_k \sum^N_{y=1} M^{(y)}_m  \rangle -\langle \sum^N_{y=1} M^{(y)}_m M^{(y)}_n \sum^N_{i=1} M^{(i)}_k \sum^N_{j=1} M^{(j)}_l  \rangle \nonumber \\
    &\; +2\langle \sum^N_{i=1} M^{(i)}_k M^{(i)}_l M^{(i)}_m \sum^N_{h=1}M^{(h)}_n \rangle +2\langle \sum^N_{j=1} M^{(j)}_l M^{(j)}_m M^{(j)}_n \sum^N_{i=1}M^{(i)}_k \rangle +2\langle \sum^N_{y=1} M^{(y)}_m M^{(y)}_n M^{(y)}_k \sum^N_{j=1}M^{(j)}_l \rangle +2\langle \sum^N_{h=1} M^{(h)}_n M^{(h)}_k M^{(h)}_l \sum^N_{y=1}M^{(y)}_m \rangle \;.
\end{align}
which gives
\begin{align}
    \mathcal{S}_{0000} 
    &= 16 \langle \hat{S}^4_{\vec{n}} \rangle -6N+3N^2-24N\langle \hat{S}^2_{\vec{n}}\rangle +32\langle \hat{S}^2_{\vec{n}}\rangle ,\\
    \mathcal{S}_{1111} &= 16 \langle \hat{S}^4_{\vec{m}} \rangle -6N+3N^2-24N\langle \hat{S}^2_{\vec{m}}\rangle +32\langle \hat{S}^2_{\vec{m}}\rangle ,\\
    \mathcal{S}_{0001} &= 4\langle \hat{S}_{\vec{n}}\hat{S}_{\vec{n}}\hat{S}_{\vec{n}}\hat{S}_{\vec{m}} +\hat{S}_{\vec{n}}\hat{S}_{\vec{n}}\hat{S}_{\vec{m}}\hat{S}_{\vec{n}} +\hat{S}_{\vec{n}}\hat{S}_{\vec{m}}\hat{S}_{\vec{n}}\hat{S}_{\vec{n}} +\hat{S}_{\vec{m}}\hat{S}_{\vec{n}}\hat{S}_{\vec{n}}\hat{S}_{\vec{n}}\rangle + (3N^2-6N)(\vec{n}\cdot \vec{m}) \nonumber\\
    & \; + (8-6N)\langle \hat{S}_{\vec{n}}\hat{S}_{\vec{m}}+\hat{S}_{\vec{m}}\hat{S}_{\vec{n}} \rangle +(16-12N)(\vec{n}\cdot \vec{m}) \langle \hat{S}^2_{\vec{n}} \rangle ,\\
    \mathcal{S}_{0111} &= 4\langle \hat{S}_{\vec{m}}\hat{S}_{\vec{m}}\hat{S}_{\vec{m}}\hat{S}_{\vec{n}} +\hat{S}_{\vec{m}}\hat{S}_{\vec{m}}\hat{S}_{\vec{n}}\hat{S}_{\vec{m}} +\hat{S}_{\vec{m}}\hat{S}_{\vec{n}}\hat{S}_{\vec{m}}\hat{S}_{\vec{m}} +\hat{S}_{\vec{n}}\hat{S}_{\vec{m}}\hat{S}_{\vec{m}}\hat{S}_{\vec{m}}\rangle + (3N^2-6N)(\vec{n}\cdot \vec{m}) \nonumber\\
    & \; + (8-6N)\langle \hat{S}_{\vec{n}}\hat{S}_{\vec{m}}+\hat{S}_{\vec{m}}\hat{S}_{\vec{n}} \rangle +(16-12N)(\vec{n}\cdot \vec{m}) \langle \hat{S}^2_{\vec{m}} \rangle ,\\
    \mathcal{S}_{0011} &= \frac{1}{6} [16 \langle \hat{S}_{\vec{n}}\hat{S}_{\vec{n}}\hat{S}_{\vec{m}}\hat{S}_{\vec{m}} +\hat{S}_{\vec{n}}\hat{S}_{\vec{m}}\hat{S}_{\vec{n}}\hat{S}_{\vec{m}} +\hat{S}_{\vec{n}}\hat{S}_{\vec{m}}\hat{S}_{\vec{m}}\hat{S}_{\vec{n}} +\hat{S}_{\vec{m}}\hat{S}_{\vec{m}}\hat{S}_{\vec{n}}\hat{S}_{\vec{n}} +\hat{S}_{\vec{m}}\hat{S}_{\vec{n}}\hat{S}_{\vec{m}}\hat{S}_{\vec{n}} +\hat{S}_{\vec{m}}\hat{S}_{\vec{n}}\hat{S}_{\vec{n}}\hat{S}_{\vec{m}} \rangle \nonumber \\
    & \; -12N +6N^2 +12N^2 (\vec{n} \cdot \vec{m})^2 -24N(\vec{n} \cdot \vec{m})^2 \nonumber \\
    & \; -24N \langle \hat{S}^2_{\vec{m}}\rangle-24N\langle \hat{S}^2_{\vec{n}}\rangle -48N(\vec{n} \cdot \vec{m})\langle \hat{S}_{\vec{n}}\hat{S}_{\vec{m}}+\hat{S}_{\vec{m}}\hat{S}_{\vec{n}} \rangle \nonumber \\
    & \; +32\langle \hat{S}^2_{\vec{m}}\rangle + 32\langle \hat{S}^2_{\vec{n}}\rangle +64(\vec{n} \cdot \vec{m})\langle \hat{S}_{\vec{n}}\hat{S}_{\vec{m}}+\hat{S}_{\vec{m}}\hat{S}_{\vec{n}} \rangle ] \;.
\end{align}

These results allows us to express straightforwardly any PIBI involving correlators of order $K\leq 4$ in terms of collective spin operator. The result is a so-called Bell correlation witness, which is an inequality of practical experimental test. For PIBIs involving correlators of order $K> 4$, expressions similar to the one presented in this section have to be found, a task which can be significantly simplified by running a computer program.

\clearpage
\newpage
\section{III.\quad Expectation values of collective spin high-order moments for OAT states}\label{si:section2}

For a system of $N$ spin-1/2 particles, the OAT Hamiltonian in terms of the collective spin operator reads $H=\hbar\chi S_z^2$. After an adimensional interaction time $\mu=2\chi t$, a coherent spin state along the $x$-direction evolves into the state
\begin{align}
    |\Phi(\mu)\rangle=\frac{1}{\sqrt{2^N}} \sum_{k=0}^N \sqrt{\binom{N}{k}} e^{-i(N/2-k)^2 \mu / 2}|k\rangle   \;,
\end{align}
where $|k\rangle$ denotes the $N$-particle Dicke state with $k$ excitations. 

Expectation values of arbitrary moments of the collective spin can be computed analytically. For a collective spin along direction $\vec{u}=(u_x,u_y,u_z)$, we find the moments up to order four to be
%
\begin{align}
    \langle \hat{S}_{\vec{u}} \rangle &= S u_x \cos^{-1+2S}{(\mu/2)} ,\\
    \langle \hat{S}_{\vec{u}}^2 \rangle &= \frac{1}{4}S\left[  (1+2S)(u_x^2+u_y^2)+2u_z^2+(-1+2S)\left((u_x-u_y)(u_x+u_y)\cos^{2(-1+S)}{(\mu)}+4u_y u_z\cos^{-2+2S}{(\mu/2)}\sin{(\mu/2)} \right)
    \right] ,\\
    \langle \hat{S}_{\vec{u}}^3 \rangle &= \frac{1}{8} S \left[  u_x\cos^{-3+2S}{(\mu/2)}\left((1-3S+6S^2)(u_x^2+u_y^2)-4(2+3(-2+S)S)u_z^2+2\left((-1+3S)(u_x^2+u_y^2) +2(1+3(-1+S)S)u_z^2 \right)\cos{(\mu)} \right) \right. \nonumber \\
    &\;\left.+ (-1+S)(-1+2S)u_x(u_x^2-3u_y^2)\cos^{-3+2S}{(3\mu/2)}+12(-1+S)(-1+2S)u_x u_y u_z \cos^{-3+2S}{(\mu)}\sin{(\mu)}
    \right] ,\\
    \langle \hat{S}_{\vec{u}}^4 \rangle &= \frac{1}{32} S \left\{
    (-1+S+12S^2+12S^3)(u_x^2+u_y^2)^2+8S(-1+6S)(u_x^2+u_y^2)u_z^2+8(-1+3S)u_z^4 \right. \nonumber\\
    &\; +(-1+2S) \left[
    (-1+S)(-3+2S)(u_x^4-6u_x^2u_y^2+u_y^4)\cos^{-4+2S}(2\mu) +4(u_x-u_y)(u_x+u_y)\cos^{-4+2S}{(\mu)} \right. \nonumber\\
    &\; \times \left(
    u_x^2-2Su_x^2+2S^2u_x^2+u_y^2-2Su_y^2+2S^2u_y^2-11u_z^2+18Su_z^2-6S^2u_z^2-2u_x^2\cos{(2\mu)}+3Su_x^2\cos{(2\mu)} \right. \nonumber\\ 
    &\; \left.-2u_y^2\cos{(2\mu)}+3Su_y^2\cos{(2\mu)}+7u_z^2\cos{(2\mu)}-12Su_z^2\cos{(2\mu)}+6S^2u_z^2\cos{(2\mu)}
    \right) \nonumber\\
    &\; -8(-1+S)(-3+2S)u_y(-3u_x^2+u_y^2)u_z \cos^{-4+2S}{(3\mu/2)}\sin{(3\mu/2)}+8u_y u_z\cos^{-4+2S}{(\mu/2)} \nonumber\\
    &\; \times \left(7u_x^2\sin{(\mu/2)}-12Su_x^2\sin{(\mu/2)}+6S^2u_x^2\sin{(\mu/2)}+7u_y^2\sin{(\mu/2)}-12Su_y^2\sin{(\mu/2)}+6S^2u_y^2\sin{(\mu/2)} \right. \nonumber\\
    &\; -11u_z^2\sin{(\mu/2)}+18Su_z^2\sin{(\mu/2)} -6S^2u_z^2\sin{(\mu/2)}-2u_x^2\sin{(3\mu/2)}+3Su_x^2\sin{(3\mu/2)} \nonumber \\
    &\;\left.\left.\left.-2u_y^2\sin{(3\mu/2)}+3Su_y^2\sin{(3\mu/2)}+u_z^2\sin{(3\mu/2)}-2Su_z^2\sin{(3\mu/2)}+2S^2u_z^2\sin{(3\mu/2)}
    \right)
    \right]
    \right\} ,
\end{align}
where $S=N/2$ is total spin.

These expressions allow us to conveniently compute the violation of Bell correlation witnesses for OAT states, as a function of the chosen measurement directions.

\clearpage
\newpage
\section{IV.\quad  Writing collective spin expectation values in terms of $\mathcal{S}$ correlators}\label{si:section2}

To construct experimentally practical Bell operators, our approach consists of starting from a set of convenient measurements, and derive from them the structure of the associated Bell inequalities.
This is enabled by the fact that expectation values of a collective spin operator can be written in terms of the permutationally invariant $\mathcal{S}$ correlators. To see this, note first that the results presented in Supplementary Sec.II can be inverted to write
\begin{align}
    & \langle \hat{S}_{\vec{n}} \rangle  = \frac{\mathcal{S}_0}{2} ,\\
    & \langle \hat{S}_{\vec{m}} \rangle  = \frac{\mathcal{S}_1}{2} ,\\
    & \langle \hat{S}^2_{\vec{n}} \rangle  = \frac{\mathcal{S}_{00}+N}{4} ,\\
    & \langle \hat{S}^2_{\vec{m}} \rangle  = \frac{\mathcal{S}_{11}+N}{4} ,\\
    & \langle \hat{S}_{\vec{n}}\hat{S}_{\vec{m}} + \hat{S}_{\vec{m}}\hat{S}_{\vec{n}} \rangle = \frac{\mathcal{S}_{01}+(\vec{n}\cdot \vec{m})N}{2} ,\\
    & \langle \hat{S}^3_{\vec{n}} \rangle  = \frac{\mathcal{S}_{000}+(3N-2)\mathcal{S}_0}{8} ,\\
    & \langle \hat{S}^3_{\vec{m}} \rangle = \frac{\mathcal{S}_{111}+(3N-2)\mathcal{S}_1}{8} ,\\
    & \langle \hat{S}_{\vec{n}}\hat{S}_{\vec{n}}\hat{S}_{\vec{m}}+ \hat{S}_{\vec{n}}\hat{S}_{\vec{m}}\hat{S}_{\vec{n}}+\hat{S}_{\vec{m}}\hat{S}_{\vec{n}}\hat{S}_{\vec{n}} \rangle = \frac{3\mathcal{S}_{001}-(2-3N)\mathcal{S}_1-(4-6N)(\vec{n}\cdot\vec{m})\mathcal{S}_0}{8} ,\\
    & \langle \hat{S}_{\vec{m}}\hat{S}_{\vec{m}}\hat{S}_{\vec{n}}+ \hat{S}_{\vec{m}}\hat{S}_{\vec{n}}\hat{S}_{\vec{m}}+\hat{S}_{\vec{n}}\hat{S}_{\vec{m}}\hat{S}_{\vec{m}} \rangle = \frac{3\mathcal{S}_{011}-(2-3N)\mathcal{S}_0-(4-6N)(\vec{n}\cdot\vec{m})\mathcal{S}_1}{8}, \\
    & \langle \hat{S}^4_{\vec{n}} \rangle = \frac{\mathcal{S}_{0000}+2(3N-4)(\mathcal{S}_{00}+N)-3(N-2)N}{16} ,\\
    & \langle \hat{S}^4_{\vec{m}} \rangle = \frac{\mathcal{S}_{1111}+2(3N-4)(\mathcal{S}_{11}+N)-3(N-2)N}{16} , \\
    & \langle \hat{S}_{\vec{n}} \hat{S}_{\vec{n}}\hat{S}_{\vec{n}}\hat{S}_{\vec{m}}+  \hat{S}_{\vec{n}} \hat{S}_{\vec{n}}\hat{S}_{\vec{m}}\hat{S}_{\vec{n}}+\hat{S}_{\vec{n}} \hat{S}_{\vec{m}}\hat{S}_{\vec{n}}\hat{S}_{\vec{n}}+\hat{S}_{\vec{m}} \hat{S}_{\vec{n}}\hat{S}_{\vec{n}}\hat{S}_{\vec{n}} \rangle \nonumber \\
    &\qquad = \frac{\mathcal{S}_{0001}-3(N-2)N(\vec{n}\cdot \vec{m})+(3N-4)(\vec{n}\cdot\vec{m})(\mathcal{S}_{00}+N)-(4-3N)(\mathcal{S}_{01}+N(\vec{n}\cdot\vec{m}))}{4} ,\\
    %
    %
    & \langle \hat{S}_{\vec{m}} \hat{S}_{\vec{m}}\hat{S}_{\vec{m}}\hat{S}_{\vec{n}}+  \hat{S}_{\vec{m}} \hat{S}_{\vec{m}}\hat{S}_{\vec{n}}\hat{S}_{\vec{m}}+\hat{S}_{\vec{m}} \hat{S}_{\vec{n}}\hat{S}_{\vec{m}}\hat{S}_{\vec{m}}+\hat{S}_{\vec{n}} \hat{S}_{\vec{m}}\hat{S}_{\vec{m}}\hat{S}_{\vec{m}} \rangle \nonumber \\
    &\qquad= \frac{\mathcal{S}_{0111}-3(N-2)N(\vec{n}\cdot \vec{m})+(3N-4)(\vec{n}\cdot\vec{m})(\mathcal{S}_{11}+N)-(4-3N)(\mathcal{S}_{01}+N(\vec{n}\cdot\vec{m}))}{4} ,\\
    & \langle \hat{S}_{\vec{n}}\hat{S}_{\vec{n}}\hat{S}_{\vec{m}}\hat{S}_{\vec{m}} +\hat{S}_{\vec{n}}\hat{S}_{\vec{m}}\hat{S}_{\vec{n}}\hat{S}_{\vec{n}}+\hat{S}_{\vec{n}}\hat{S}_{\vec{m}}\hat{S}_{\vec{m}}\hat{S}_{\vec{n}}+\hat{S}_{\vec{m}}\hat{S}_{\vec{m}}\hat{S}_{\vec{n}}\hat{S}_{\vec{n}} +\hat{S}_{\vec{m}}\hat{S}_{\vec{n}}\hat{S}_{\vec{m}}\hat{S}_{\vec{n}} +\hat{S}_{\vec{m}}\hat{S}_{\vec{n}}\hat{S}_{\vec{n}}\hat{S}_{\vec{m}} \rangle \nonumber \\
    &\qquad = \frac{3\mathcal{S}_{0011}+(3N-4)(\mathcal{S}_{00}+\mathcal{S}_{11}+4(\vec{n}\cdot\vec{m})\mathcal{S}_{01})+N(3N-2)\left(2(\vec{n}\cdot\vec{m})^2+1 \right)}{8} \;.
\end{align}
Using the (linearly independent) measurement direction $\vec{m}$ and $\vec{n}$, we can write any measurement direction $\vec{a}$ lying on the $\{\vec{m},\vec{n}\}$ plane as
\begin{align}
\vec{a}=\alpha \vec{m} +\beta \vec{n}
\end{align}
with normalized weights $|\alpha|^2+|\beta|^2=1$, and $ \alpha,\beta\in\mathbb{R}^3$.

Expectation values for the first four moments of the collective spin operator along some direction $\vec{a}$ are thus

\begin{align}
    \langle \hat{S}_{\vec{a}} \rangle &= \alpha \langle \hat{S}_{\vec{m}} \rangle + \beta \langle \hat{S}_{\vec{n}} \rangle \nonumber \\
     &= \frac{\beta}{2} \mathcal{S}_0 + \frac{\alpha}{2} \mathcal{S}_1 , \\
    \langle \hat{S}^2_{\vec{a}} \rangle &= \alpha^2 \langle \hat{S}^2_{\vec{m}} \rangle + \beta^2 \langle \hat{S}^2_{\vec{n}} \rangle + \alpha \beta \langle \hat{S}_{\vec{m}}\hat{S}_{\vec{n}} + \hat{S}_{\vec{n}} \hat{S}_{\vec{m}} \rangle \nonumber \\
    &= \frac{\beta^2}{4}\mathcal{S}_{00} + \frac{\alpha^2}{4} \mathcal{S}_{11} + \frac{\alpha\beta}{2}\mathcal{S}_{01} + \left( \alpha^2+\beta^2+2\alpha\beta(\vec{n}\cdot\vec{m}) \right) \frac{N}{4} ,\\
    \langle \hat{S}^3_{\vec{a}} \rangle &= \alpha^3 \langle \hat{S}^3_{\vec{m}} \rangle + \beta^3 \langle \hat{S}^3_{\vec{n}} \rangle + \alpha^2 \beta \langle \hat{S}_{\vec{m}} \hat{S}_{\vec{m}} \hat{S}_{\vec{n}} + \hat{S}_{\vec{m}} \hat{S}_{\vec{n}} \hat{S}_{\vec{m}} + \hat{S}_{\vec{n}} \hat{S}_{\vec{m}} \hat{S}_{\vec{m}} \rangle +\alpha \beta^2 \langle \hat{S}_{\vec{n}} \hat{S}_{\vec{n}} \hat{S}_{\vec{m}} + \hat{S}_{\vec{n}} \hat{S}_{\vec{m}} \hat{S}_{\vec{n}} + \hat{S}_{\vec{m}} \hat{S}_{\vec{n}} \hat{S}_{\vec{n}} \rangle \nonumber \\
    &= \frac{\beta^3}{8}\mathcal{S}_{000} + \frac{\alpha^3}{8}\mathcal{S}_{111} + \frac{3\alpha \beta^2}{8}\mathcal{S}_{001} + \frac{3\alpha^2\beta}{8} \mathcal{S}_{011}  \nonumber  \\
    &\;+\frac{(3N-2)}{8} \left(\beta^3+\alpha^2\beta+2\alpha\beta^2(\vec{n}\cdot\vec{m}) \right) \mathcal{S}_0 + \frac{(3N-2)}{8} \left(\alpha^3+\alpha\beta^2+2\alpha^2\beta(\vec{n}\cdot\vec{m}) \right)\mathcal{S}_1 \\
    \langle \hat{S}_{\vec{a}}^4 \rangle &= \alpha^4 \langle \hat{S}^4_{\vec{m}} \rangle + \beta^4 \langle \hat{S}^4_{\vec{n}} \rangle + \alpha^3\beta \langle \hat{S}_{\vec{m}}\hat{S}_{\vec{m}}\hat{S}_{\vec{m}}\hat{S}_{\vec{n}} + \hat{S}_{\vec{m}}\hat{S}_{\vec{m}}\hat{S}_{\vec{n}}\hat{S}_{\vec{m}} + \hat{S}_{\vec{m}}\hat{S}_{\vec{n}}\hat{S}_{\vec{m}}\hat{S}_{\vec{m}} + \hat{S}_{\vec{n}}\hat{S}_{\vec{m}}\hat{S}_{\vec{m}}\hat{S}_{\vec{m}} \rangle \nonumber \\
    & \;+ \alpha\beta^3 \langle \hat{S}_{\vec{n}}\hat{S}_{\vec{n}}\hat{S}_{\vec{n}}\hat{S}_{\vec{m}} + \hat{S}_{\vec{n}}\hat{S}_{\vec{n}}\hat{S}_{\vec{m}}\hat{S}_{\vec{n}} + \hat{S}_{\vec{n}}\hat{S}_{\vec{m}}\hat{S}_{\vec{n}}\hat{S}_{\vec{n}} + \hat{S}_{\vec{m}}\hat{S}_{\vec{n}}\hat{S}_{\vec{n}}\hat{S}_{\vec{n}} \rangle \nonumber \\
    &\;+ \alpha^2\beta^2 \langle \hat{S}_{\vec{m}}\hat{S}_{\vec{m}}\hat{S}_{\vec{n}}\hat{S}_{\vec{n}} + \hat{S}_{\vec{m}}\hat{S}_{\vec{n}}\hat{S}_{\vec{m}}\hat{S}_{\vec{n}} +\hat{S}_{\vec{m}}\hat{S}_{\vec{n}}\hat{S}_{\vec{n}}\hat{S}_{\vec{m}} + \hat{S}_{\vec{n}}\hat{S}_{\vec{n}}\hat{S}_{\vec{m}}\hat{S}_{\vec{m}} + \hat{S}_{\vec{n}}\hat{S}_{\vec{m}}\hat{S}_{\vec{n}}\hat{S}_{\vec{m}} +\hat{S}_{\vec{n}}\hat{S}_{\vec{m}}\hat{S}_{\vec{m}}\hat{S}_{\vec{n}} \rangle \nonumber \\
    &= \frac{\beta^4}{16} \mathcal{S}_{0000} +\frac{\alpha^4}{16} \mathcal{S}_{1111} +\frac{\alpha\beta^3}{4}\mathcal{S}_{0001}+\frac{\alpha^3\beta}{4}\mathcal{S}_{0111}+\frac{3\alpha^2\beta^2}{8} \mathcal{S}_{0011} \nonumber \\
    &\;+(3N-4)\left(\frac{\beta^4}{8}+\frac{\alpha\beta^3}{4}(\vec{n}\cdot\vec{m})+\frac{\alpha^2\beta^2}{8} \right) \mathcal{S}_{00} \nonumber \\
    &\;+(3N-4)\left(\frac{\alpha^4}{8}+\frac{\alpha^3\beta}{4}(\vec{n}\cdot\vec{m})+\frac{\alpha^2\beta^2}{8} \right) \mathcal{S}_{11} \nonumber \\
    &\;+(3N-4)\left(\frac{\alpha^3\beta}{4}+\frac{\alpha\beta^3}{4}+\frac{\alpha^2\beta^2}{2}(\vec{n}\cdot\vec{m}) \right) \mathcal{S}_{01} \nonumber \\
    &\;+N(3N-2) \left(\frac{\alpha^4}{16}+\frac{\beta^4}{16}+\frac{\alpha^3\beta}{4}(\vec{n}\cdot\vec{m})+\frac{\alpha\beta^3}{4}(\vec{n}\cdot\vec{m})+\frac{\alpha^2\beta^2}{8}(2(\vec{n}\cdot\vec{m})^2+1) \right) . \label{eq:Sa4}
\end{align}
These results allow us to derive the form of the PIBI associated to a specific Bell operator.

\clearpage
\newpage
\section{V.\quad Families of PIBI\lowercase{s} involving third-order correlators}

Besides the PIBI $I_3$ presented in the main text, we also find the following PIBIs involving third-order correlators. See Fig.~\ref{SI_Fig4} for a plot of their relative quantum violation as a function of $N$.
\begin{align}
    I_3^{(1)} &\equiv -24(N-1)\mathcal{S}_0+3(N+2)\mathcal{S}_{00}-6(N-2)\mathcal{S}_{01}+3(N-2)\mathcal{S}_{11}-2\mathcal{S}_{000}+3\mathcal{S}_{001}-\mathcal{S}_{111}+12N(N-1) \geq 0 , \nonumber \\
    I_3^{(2)} &\equiv -4(3N-1)\mathcal{S}_0-4(N-1)\mathcal{S}_1+(N+6)\mathcal{S}_{00}+2N\mathcal{S}_{01}+(N-2)\mathcal{S}_{11}-\mathcal{S}_{000}-2\mathcal{S}_{001}-\mathcal{S}_{011}+4N(N+1) \geq 0 , \nonumber\\
    I_3^{(3)} &\equiv -8N\mathcal{S}_0-8(N-1)\mathcal{S}_1+(N+2)\mathcal{S}_{00}+2(N+2)\mathcal{S}_{01}+(N-2)\mathcal{S}_{11}-\mathcal{S}_{000}-2\mathcal{S}_{001}-\mathcal{S}_{011}+4N(N+1) \geq 0 , \nonumber\\
    I_3^{(4)} &\equiv -8(N-1)\mathcal{S}_0-8\mathcal{S}_1+(N-2)\mathcal{S}_{00}-2(N-6)\mathcal{S}_{01}+(N-6)\mathcal{S}_{11}-\mathcal{S}_{001}+2\mathcal{S}_{011}-\mathcal{S}_{111}+4N(N+1) \geq 0 , \nonumber\\
    I_3^{(5)} &\equiv -8(N-1)\mathcal{S}_0+(N+2)\mathcal{S}_{00}-2(N-2)\mathcal{S}_{01}+(N-2)\mathcal{S}_{11}-\mathcal{S}_{000}+2\mathcal{S}_{001}-\mathcal{S}_{011}+4N(N-1) \geq 0 , \nonumber\\
    I_3^{(6)} &\equiv -4(N-1)\mathcal{S}_0-4(N-1)\mathcal{S}_1+(N-2)\mathcal{S}_{00}-2(N-4)\mathcal{S}_{01}+(N-2)\mathcal{S}_{11}-\mathcal{S}_{000}+2\mathcal{S}_{001}-\mathcal{S}_{011}+4N(N-1) \geq 0 , \nonumber\\
    I_3^{(7)} &\equiv -4(N-1)\mathcal{S}_0-4(N-1)\mathcal{S}_1+(N-2)\mathcal{S}_{00}+2N\mathcal{S}_{01}+(N-2)\mathcal{S}_{11}-\mathcal{S}_{000}-2\mathcal{S}_{001}-\mathcal{S}_{011}+4N(N-1) \geq 0 , \nonumber\\
    I_3^{(8)} &\equiv (N-6)\mathcal{S}_{00}-2(N-2)\mathcal{S}_{01}+(N-2)\mathcal{S}_{11}-\mathcal{S}_{000}+2\mathcal{S}_{001}-\mathcal{S}_{011}+4N(N-1) \geq 0 ,  \nonumber\\
    I_3^{(9)} &\equiv 3(N-4)\mathcal{S}_{00}-6N\mathcal{S}_{01}+3N\mathcal{S}_{11}-2\mathcal{S}_{000}+3\mathcal{S}_{001}-\mathcal{S}_{111}+12N(N-1) \geq 0 , \nonumber\\
    I_3^{(10)} &\equiv -12(N-1)\mathcal{S}_0-4(N-7)\mathcal{S}_1+(N-2)\mathcal{S}_{00}+2(N-8)\mathcal{S}_{01}+(N-10)\mathcal{S}_{11}+\mathcal{S}_{001}+2\mathcal{S}_{011}+\mathcal{S}_{111}+4N(N+5) \geq 0 , \nonumber\\
    I_3^{(11)} &\equiv -12(N-1)\mathcal{S}_0+12(N+1)\mathcal{S}_1+(N-2)\mathcal{S}_{00}-2(N+4)\mathcal{S}_{01}+(N+6)\mathcal{S}_{11}+\mathcal{S}_{001}-2\mathcal{S}_{011}+\mathcal{S}_{111}+4N(N+5) \geq 0 , \nonumber\\
    I_3^{(12)} &\equiv 8(N-1)\mathcal{S}_0+8(2N+1)\mathcal{S}_1+(N-2)\mathcal{S}_{00}+2(N+2)\mathcal{S}_{01}+(N+10)\mathcal{S}_{11}+\mathcal{S}_{001}+2\mathcal{S}_{011}+\mathcal{S}_{111}+4N(N+5) \geq 0 , \nonumber\\
    I_3^{(13)} &\equiv -16(N-1)\mathcal{S}_0-8(N-7)\mathcal{S}_1+(N-2)\mathcal{S}_{00}+2(N-10)\mathcal{S}_{01}+(N-14)\mathcal{S}_{11}+\mathcal{S}_{001}+2\mathcal{S}_{011}+\mathcal{S}_{111}+4N(N+11) \geq 0 , \nonumber\\
    I_3^{(14)} &\equiv -16(N-1)\mathcal{S}_0+16(N+2)\mathcal{S}_1+(N-2)\mathcal{S}_{00}-2(N+6)\mathcal{S}_{01}+(N+10)\mathcal{S}_{11}+\mathcal{S}_{001}-2\mathcal{S}_{011}+\mathcal{S}_{111}+4N(N+11) \geq 0 , \nonumber\\
    I_3^{(15)} &\equiv 12(N-1)\mathcal{S}_0+4(5N+7)\mathcal{S}_1+(N-2)\mathcal{S}_{00}+2(N+4)\mathcal{S}_{01}+(N+14)\mathcal{S}_{11}+\mathcal{S}_{001}+2\mathcal{S}_{011}+\mathcal{S}_{111}+4N(N+11) \geq 0 , \nonumber\\
    I_3^{(16)} &\equiv (20N-44)\mathcal{S}_0+(2N+2)\mathcal{S}_{00}+(-4N+16)\mathcal{S}_{01}+(2N-8)\mathcal{S}_{11}+\mathcal{S}_{000}-\mathcal{S}_{001}-\mathcal{S}_{011}+\mathcal{S}_{111}+(10N^2-34N+48) \geq 0 , \nonumber\\
    I_3^{(17)} &\equiv 24(N-3)\mathcal{S}_0+24(N-3)\mathcal{S}_1+5(N-2)\mathcal{S}_{00}-2(5N-22)\mathcal{S}_{01}+5(N-2)\mathcal{S}_{11} +2\mathcal{S}_{000}-\mathcal{S}_{001}-4\mathcal{S}_{011}+3\mathcal{S}_{111} \nonumber \\
    &\;+24(N^2-5N+8) \geq 0 .
\end{align}
\begin{figure}[t]
	\begin{center}
		\includegraphics[width=110mm]{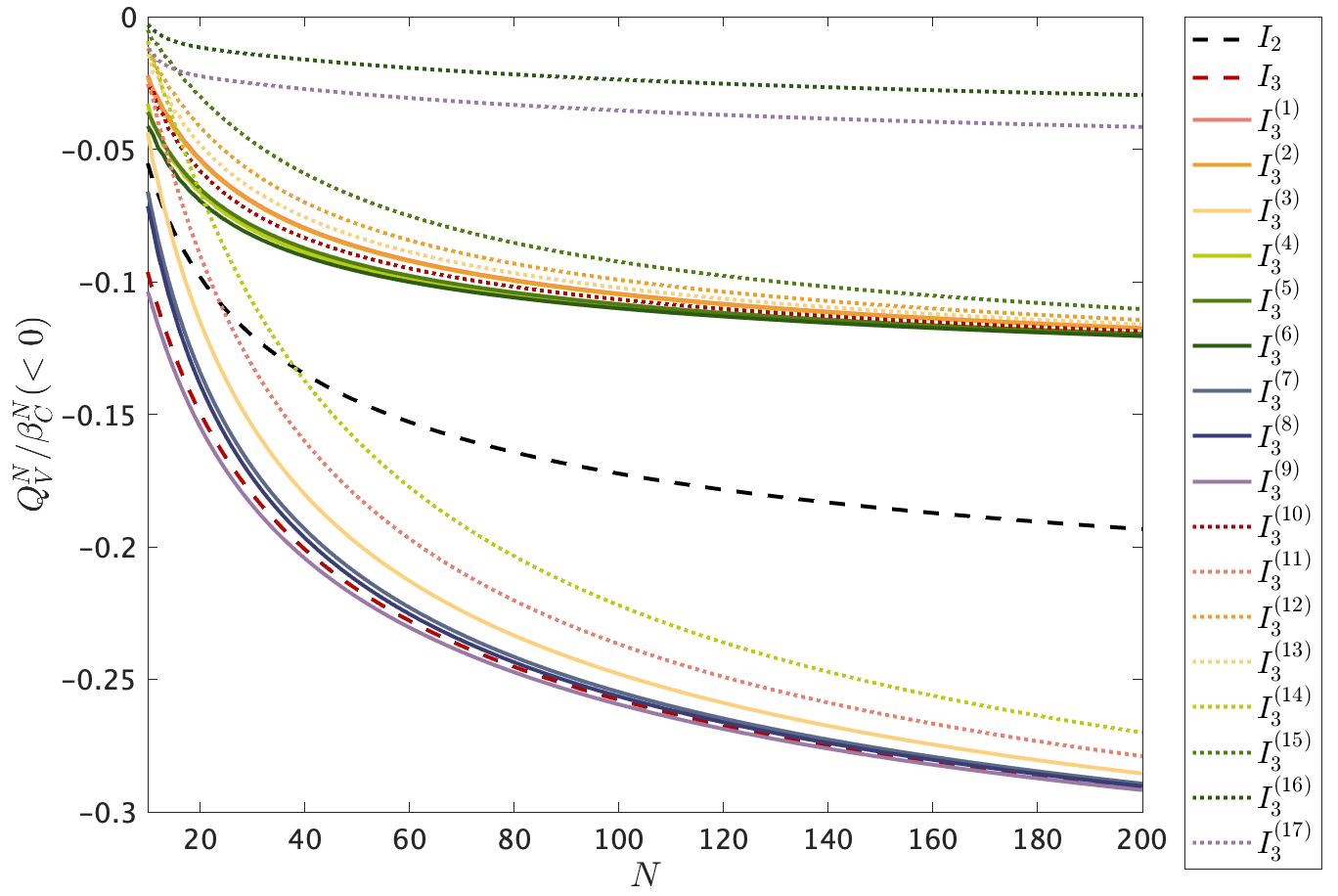}
	\end{center}
	\caption{Maximum relative quantum violation of the PIBIs $I_3^{(i)}$, compared to $I_3$ and $I_2$ of our main text. }
	\label{SI_Fig4}
\end{figure}
%

\newpage
\section{VI.\quad PIBI involving fourth-order correlators}

If we only take into account the second and the fourth order correlators, by finding the corresponding local polytope we find the PIBI
\begin{align}
    I_4=24(N-1)\mathcal{S}_{00}+48(N-1)\mathcal{S}_{01}+24(N-3)\mathcal{S}_{11}+\mathcal{S}_{0000}+4\mathcal{S}_{0001}+6\mathcal{S}_{0011}+4\mathcal{S}_{0111}+\mathcal{S}_{1111}+48N(N-1)\geq 0 .
\end{align}
In Fig.~\ref{SI_Fig6} we compare its maximum relative quantum violation to the one of $I_2$ and $I_3$ of the main text, as a function of $N$.
\begin{figure}[t]
	\begin{center}
		\includegraphics[width=110mm]{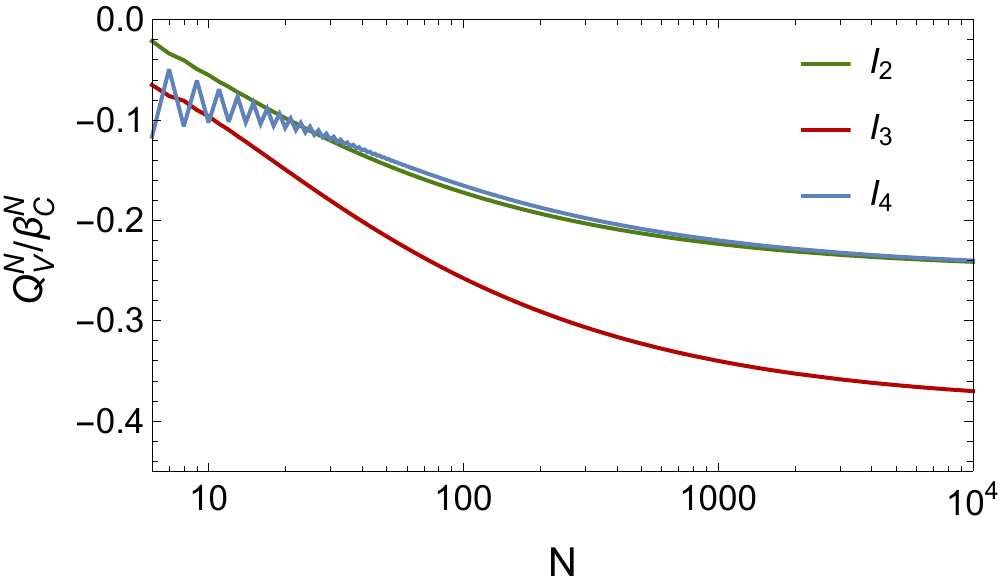}
	\end{center}
	\caption{Maximum relative quantum violation of the fourth-order PIBI $I_4$, compared to $I_3$ and $I_2$ of our main text.}
	\label{SI_Fig6}
\end{figure}

According to Eq.~(\ref{eq:Sa4}), when $\alpha=\beta=\frac{1}{\sqrt{2}}$ we have $\vec{a}=(\vec{n}+\vec{m})/\sqrt{2}$ and
\begin{align}
&\mathcal{S}_{0000}+4\mathcal{S}_{0001}+6\mathcal{S}_{0011}+4\mathcal{S}_{0111}+\mathcal{S}_{1111} \nonumber \\
&= 64 \langle \hat{S}^4_{\vec{a}} \rangle -4(3N-4)(1+(\vec{n}\cdot\vec{m}))(\mathcal{S}_{00}+\mathcal{S}_{11}+2\mathcal{S}_{01}) -N(3N-2)\left[2+8(\vec{n}\cdot\vec{m})+2\left(2(\vec{n}\cdot\vec{m})^2+1 \right) \right]
\end{align}
From this, we can rewrite $I_4$ as
\begin{align}
    I_4 
    &= 64\langle \hat{S}^4_{\vec{a}} \rangle +\left[-4(3N-4)(\vec{n}\cdot\vec{m})+4(3N-2) \right]\mathcal{S}_{00} +\left[-4(3N-4)(\vec{n}\cdot\vec{m})+4(3N-14) \right] \mathcal{S}_{11} \nonumber \\
    &\; +\left[-8(3N-4)(\vec{n}\cdot\vec{m})+8(3N-2) \right]\mathcal{S}_{01} +\left[-N(3N-2)\left[2+8(\vec{n}\cdot\vec{m})+2\left(2(\vec{n}\cdot\vec{m})^2+1 \right) \right]+48N(N-1)\right] \nonumber \\
    &= 64\langle \hat{S}^4_{\vec{a}} \rangle -192 \langle \hat{S}^2_{\vec{m}} \rangle +32[-\sqrt{2}(3N-4)(\vec{a}\cdot\vec{m})+6(N-1)]\langle \hat{S}^2_{\vec{a}} \rangle +12(\sqrt{2}(\vec{a}\cdot\vec{m})-2)\left[\sqrt{2}(\vec{a}\cdot\vec{m})(N-2)-2N\right]N
\end{align}
which shows that only two measurement directions, namely $\vec{m}$ and $\vec{a}$, are needed to test $I_4$ experimentally.

\clearpage
\newpage
\section{VII.\quad SDP method}

In an $N$-party scenario with two measurement settings per party and two possible outputs per measurement, there exist four local deterministic strategies (LDSs) per party. For the characterization of permutationally invariant inequalities, the vertices of the local polytope are naturally projected onto the symmetric space via permutation of the correlators. Such a projection onto the symmetric space induces a natural classification of the local deterministic strategies in terms of the partitions of $N$ in four elements, which we denote $\{a,b,c,d\}$, corresponding to Eq.~(\ref{eq:abcd}). These represent how many parties corresponding to each strategy. Hence, $a$, $b$, $c$, and $d$ are non-negative integers and the sum of them is $a+b+c+d=N$. $a$ can be thought of as the amount of parties that output the first outcome on the first measurement setting and the first outcome on the second measurement setting. $b$ the number of parties that output the first and second outcomes on the first and second measurement settings, respectively, and so on.

In the case where at most third-order correlators are involved, the correlator ensemble is $\vec{\mathcal{S}}_3 \equiv (\mathcal{S}_{0},\mathcal{S}_{1},\mathcal{S}_{00},\mathcal{S}_{01},\mathcal{S}_{1},\mathcal{S}_{000},\mathcal{S}_{001},\mathcal{S}_{011},\mathcal{S}_{111})$. Furthermore, since correlation functions factorize at each LDS \cite{FinePRL1982}, this factorization induces another factorization at the level of symmetric correlators, when evaluated at projected vertices \cite{AnnPhys, FadelPRL2017}. For example, it is immediate to see that ${\cal S}_{00} = {\cal S}_0^2-N$ or that ${\cal S}_{01} = {\cal S}_0 {\cal S}_1 - {\cal Z}$, where ${\cal Z} = a-b-c+d$. This allows us to express any correlator in terms of linear polynomials in $a,b,c,d$. More precisely, let us consider the following transformation
\begin{equation}
    \left(
    \begin{array}{c}
         N\\
         {\cal S}_1\\
         {\cal S}_0\\
         {\cal Z}
    \end{array}
    \right) = 2 H^{\otimes 2}\left(
    \begin{array}{c}
         a\\
         b\\
         c\\
         d
    \end{array}
    \right) = 
    \left(
    \begin{array}{cccc}
        1 & 1&1&1 \\
        1 & -1&1&-1 \\
        1 & 1&-1&-1 \\
        1 & -1&-1&1
    \end{array}
    \right)
    \left(
    \begin{array}{c}
         a\\
         b\\
         c\\
         d
    \end{array}
    \right),
\end{equation}
where $H$ is the Hadamard gate.
However, the description in terms of the explicit parameters $a,b,c,d$ is not totally convenient for our purposes, nor is the one in terms of $N, {\cal S}_1,{\cal S}_0,{\cal Z}$. The projected symmetric polytope is embedded in a nine-dimensional affine space and, if $a,b,c,d$ were real, continuous parameters, these would form a four-dimensional manifold in that space. The natural coordinates for the projected symmetric polytope are therefore the symmetric correlators, and $N$ and ${\cal Z}$ are none of them, but just artifacts of the parameterization. However, we can relate them to coordinate axis in the polytope affine space, since $N={\cal S}_0^2-{\cal S}_{00}$ and ${\cal Z} = {\cal S}_0{\cal S}_1 - {\cal S}_{01}$.

By inverting the above relations, we can transform the inequality constraints $a,b,c,d \geq 0$ into the following respective polynomial inequalities $g_1,\ldots g_4$ in the variables of $\vec{\mathcal{S}}_3$:
 \begin{align}
 \left( \begin{matrix}
 g_1 \\
 g_2 \\
 g_3 \\
 g_4
 \end{matrix}\right) 
  = \frac{1}{4}
 \left( \begin{matrix}
  (\mathcal{S}_0^2-\mathcal{S}_{00})+\mathcal{S}_0+\mathcal{S}_1+(\mathcal{S}_0\mathcal{S}_1-\mathcal{S}_{01}) \\
  (\mathcal{S}_0^2-\mathcal{S}_{00})+\mathcal{S}_0-\mathcal{S}_1-(\mathcal{S}_0\mathcal{S}_1-\mathcal{S}_{01}) \\
  (\mathcal{S}_0^2-\mathcal{S}_{00})-\mathcal{S}_0+\mathcal{S}_1-(\mathcal{S}_0\mathcal{S}_1-\mathcal{S}_{01}) \\
  (\mathcal{S}_0^2-\mathcal{S}_{00})-\mathcal{S}_0-\mathcal{S}_1+(\mathcal{S}_0\mathcal{S}_1-\mathcal{S}_{01}) 
 \end{matrix}\right) 
 \geq 0,
 \end{align}
 where $g_i$ is shorthand of $g_i(\vec{\mathcal{S}}_3)$. 

At this stage, we have the tools to take the next step for approximating the local polytope with a continuous manifold. We first consider $\{a,b,c,d\}\in\mathbb{R}_{\geq0}$ instead of $\{a,b,c,d\}\in\mathbb{Z}_{\geq0}$, and relabel them as non-negative constraints in terms of 
 correlators $g_i(\vec{\mathcal{S}}_3)\geq0$.
These give a so-called semialgebraic set in the $\vec{\mathcal{S}}_3$ coordinates. Semialgebraic sets are defined by polynomial equality and inequality constraints and they are central objects in convex optimization \cite{ParriloBook2013}. For our purposes, the constraints are the polynomial inequalities $g_i(\vec{\mathcal{S}}_3) \geq 0$, together with the polynomial equalities such as ${\cal S}_{000} = {\cal S}_0^3-(3(\mathcal{S}_0^2-\mathcal{S}_{00})-2){\cal S}_0$, ${\mathcal S}_{001} = \mathcal{S}_0\mathcal{S}_0\mathcal{S}_1 +2\mathcal{S}_1 -(\mathcal{S}_0^2-\mathcal{S}_{00})\mathcal{S}_1-2(\mathcal{S}_0\mathcal{S}_1-\mathcal{S}_{01})\mathcal{S}_0$, etc. (cf. Eqs. (\ref{eq:S000})-(\ref{eq:S011})).

The characterization of the convex hull of a discrete set of points $\{\vec{\mathcal{S}}_3: a,b,c,d \in {\mathbb Z}, a,b,c,d \geq 0, a+b+c+d=N\}$ is now relaxed to the characterization of the convex hull of the semialgebraic set $\{\vec{\mathcal{S}}_3: a,b,c,d \in {\mathbb R}, a,b,c,d \geq 0, a+b+c+d=N\}$, which we denote $\overline{CH}({\mathbb S})$. By construction, the latter contains the former. The membership problem in the convex hull of a semialgebraic set is a subject of intense research in optimization \cite{ParriloBook2013, Gouveia2010, Gouveia2012}. Since convex hulls are intersections of half-spaces, determined by linear inequalities, the hardness of the problem lies in proving that a linear polynomial is non-negative when evaluated on a semialgebraic set. However, the question in its full generality is too complex, so a second relaxation is needed.

The second relaxation is to express the polynomial as a sum of squares so that the membership problem can be efficiently solved via SDP method \cite{Gouveia2010, Gouveia2012}. However, the sum of squares has to take into account the polynomial equality and inequality constraints.

For the equality constraints we introduce the ideal $I$ generated by a set of polynomials $f_i$
\begin{align}
I= \left\{\sum_if_ip_i\; \text{s.t.} \; p_i\in\mathbb{R}[\vec{\mathcal{S}}_3] \right\} \subseteq\mathbb{R}[\vec{\mathcal{S}}_3],
\end{align}
where $f_i=f_i(\vec{\mathcal{S}}_3)=0$ is a set of equations associated to the correlators (such as ${\cal S}_{000} - {\cal S}_0^3+(3(\mathcal{S}_0^2-\mathcal{S}_{00})-2){\cal S}_0$). Since the $f_i$ evaluate to zero in the semialgebraic set, every element in $I$ also evaluates to zero. This allows us to tremendously increase expressivity of the linear polynomials we want to show define half-spaces containing the semialgebraic set. In particular, $f_i$ allow to change the degree of such polynomials, allowing for sum-of-squares polynomials to be equivalent to e.g. linear ones when evaluated on that set.

For the inequality constraints, we proceed in a similar way: we can use any of the $g_i$ as non-negative elements in our certificate construction beacause they are, by construction, non-negative on the semialgebraic set to characterize.

All in all, this allows us to prove that a linear polynomial $l(\vec{\mathcal{S}}_3) \geq 0$ on $\overline{CH}({\mathbb S})$ via the following strategy: If we can express
\begin{equation}
    l(\vec{\mathcal{S}}_3) = \sum_{i} g_i(\vec{\mathcal{S}}_3) \sigma_i(\vec{\mathcal{S}}_3) + \sigma_0(\vec{\mathcal{S}}_3) \mod I,
    \label{eq:SoSModuloI}
\end{equation}
where $\sigma_i(\vec{\mathcal{S}}_3)$ are sum-of-squares polynomials modulo the ideal $I$ for $0\leq i \leq 4$ (i.e., there exits a set of polynomials $p_j$ such that $\sigma_i + \sum_j p_j f_j$ is a sum-of-squares), then this is a syntactic proof that every element in $\overline{CH}({\mathbb S})$ satisfies $l\geq 0$.

The condition in Eq. (\ref{eq:SoSModuloI}) can be efficiently computed, if the degree of $p_j$ is bounded, yielding a hierarchy of relaxations. Clearly, the higher the degree allowed for $p_j$, the more possibilities one can express a particular $l$ as a sum-of-squares modulo $I$, but also the higher the computational cost. The membership problem in $\overline{CH}({\mathbb S})$ can be then relaxed through a suitable modification Lasserre's moment method \cite{LasserreSIAM2001, Gouveia2010, FadelPRL2017}.

We focus on the very first level of that hierarchy (both on $\sigma_0$ and $\sigma_i$). To this end, we construct a $10\times10$ moment matrix $\Gamma_i=g_i\vec{b}\cdot\vec{b}^T$ mod $I$, with the vector $\vec{b}=(1,\mathcal{S}_0,\cdots,\mathcal{S}_{111})^T$ and linearize its components. Since we have four inequality constraints also to treat at the first level, this yields a $50\times 50$ block-diagonal matrix $\tilde{\Gamma}=\displaystyle\bigoplus_{i=0}^4 \Gamma_i$. Now, some entries of $\tilde{\Gamma}$ correspond to the correlators that are measurable in the experiment, yielding a data point denoted 
$\vec{\mathcal{S}}_3^*$. If the semidefinite program
\begin{equation}
    \begin{array}{cccc}
        \min_{\tilde{\Gamma}} &0&\\
        \mathrm{s.t.}&\tilde{\Gamma} &\succeq &0\\
        &\tilde{\Gamma}_{00} &= &1\\
        &\tilde{\Gamma}_{0i} &=&(\vec{\mathcal{S}}_3^*)_i\\
        &\tilde{\Gamma}_{ij} &=&h(\tilde{\Gamma}),
    \end{array}
    \label{eq:SdP}
\end{equation}
where $h$ encodes the constraints imposed by the ideal $I$. If the SDP (\ref{eq:SdP}) is feasible, then $\vec{\mathcal{S}}_3^*$ belongs to the spectrahedron defined by the feasible set of SDP (\ref{eq:SdP}). In that case, the SDP does not decide, the correlations $\vec{\mathcal{S}}_3^*$ might admit a LHVM or be nonlocal. However, if the SDP (\ref{eq:SdP}) is not feasible, its dual yields a certificate of infeasibility. Geometrically speaking, that is a hyperplane separating the outer approximation of $\overline{CH}({\mathbb S})$ given by the spectrahedron defined by the feasible set of the SDP (\ref{eq:SdP}) and the exterior point $\vec{\mathcal{S}}_3^*$. Since the spectrahedron contains $\overline{CH}({\mathbb S})$, infeasibility of SDP (\ref{eq:SdP}) implies that $\vec{\mathcal{S}}_3^*$ is nonlocal. Interestingly, the dual variables associated to the equality constraints $\tilde{\Gamma}_{0i} =(\vec{\mathcal{S}}_3^*)_i$ yield in that case the associated coefficients of the symmetric 3-body Bell inequality that detects $(\vec{\mathcal{S}}_3^*)_i$, the dual variable associated to the constraint $\tilde{\Gamma}_{00} = 1$ its classical bound and the dual variable associated to $\tilde{\Gamma} \succeq 0$ the sum of squares modulo $I$ of the form (\ref{eq:SoSModuloI}).

\clearpage
\newpage
\section{VIII.\quad Quantitative analysis of high-order PIBI violation and non-Gaussian properties}

To quantitatively investigate the advantage of high-order PIBIs on revealing Bell nonlocality of non-Gaussian states, in this section we investigate the relation between PIBIs violation and some measure of spin state non-Gaussian properties.

\vspace{2mm}
First, we look at the excess kurtosis, which is commonly used as a quantification for non-Gaussianity in continuous-variable systems~\cite{MattiaPRL2018,YoungNP2020}. For spin states, this quantity can be obtained by optimizing the measurement direction $\vec{a}=(\phi,\theta)$ of the collective spin $\hat{S}_{\vec{a}}$, namely
\begin{align}
K_{\text{ex}}=\min_{\phi,\theta} \left(\frac{\langle \left(\hat{S}_{\vec{a}}-\langle \hat{S}_{\vec{a}} \rangle \right)^4\rangle}{\langle \left( \hat{S}_{\vec{a}}-\langle \hat{S}_{\vec{a}}\rangle \right)^2 \rangle^2}-3 \right) \;,
\end{align}
where $K_{\text{ex}}<0$ indicates non-Gaussianity of the state.

\vspace{2mm}
In addition, there are classes of non-Gaussian quantum states that are associated to Wigner functions with negative regions. For this reason, we also look at the state's Wigner negativity, which is defined as the double volume of the negative part of the Wigner function. For spin states this is~\cite{DavidPRR2021}
\begin{equation}\label{eqWN}
    \mathcal{N} = \frac{1}{2} \left(\frac{2j+1}{4\pi} \int_{\theta=0}^{\pi}\int_{\phi=0}^{2\pi} |W_{\rho}(\theta,\phi)|\sin\theta d\theta d\phi-1 \right) \;,
\end{equation}
where $W_{\rho}(\theta,\phi)$ is the Wigner function value at point $(\theta,\phi)$ on the generalized Bloch sphere, and $j=N/2$ is the total spin. $\mathcal{N}>0$ is an indicator of non-classicality~\cite{AnatoleJOB2004}, and it has also been proved to be a necessary ingredient in quantum computation to outperform classical devices~\cite{MariPRL2012}.

\vspace{2mm}
In Figure~\ref{SI_nonGaussianity} we show for OAT states a comparison between the range over which $I_2$ and $I_3$ are violated, and the level of non-Gaussian properties of the states. Fig.~\ref{SI_nonGaussianity}(a) is a zoom of Fig.~\ref{Fig2new} in the main text, where we compare relative quantum violation between PIBI $I_2$ and $I_3$ with the increase of OAT interaction $\mu$. The dashed vertical lines indicate the critical values of $\mu$ where the PIBI violation vanishes. Non-Gaussian properties of OAT states as a function of $\mu$ are illustrated by excess kurtosis in Fig.~\ref{SI_nonGaussianity}(b) and Wigner negativity in Fig.~\ref{SI_nonGaussianity}(c), where it is evident an increase of non-Gaussianity and non-classicality with $\mu$. By looking at the vertical dashed lines it is evident that $I_3$ outperforms $I_2$ in nonlocality detection, and especially that $I_3$ allows us to detect nonlocality in states with higher levels of non-Gaussianity. This is expected from the fact that increasing the order of the correlators involved in the PIBI allows for detecting correlations that do not manifest in lower order correlators.

\begin{figure}[t]
	\begin{center}
	\includegraphics[width=70mm]{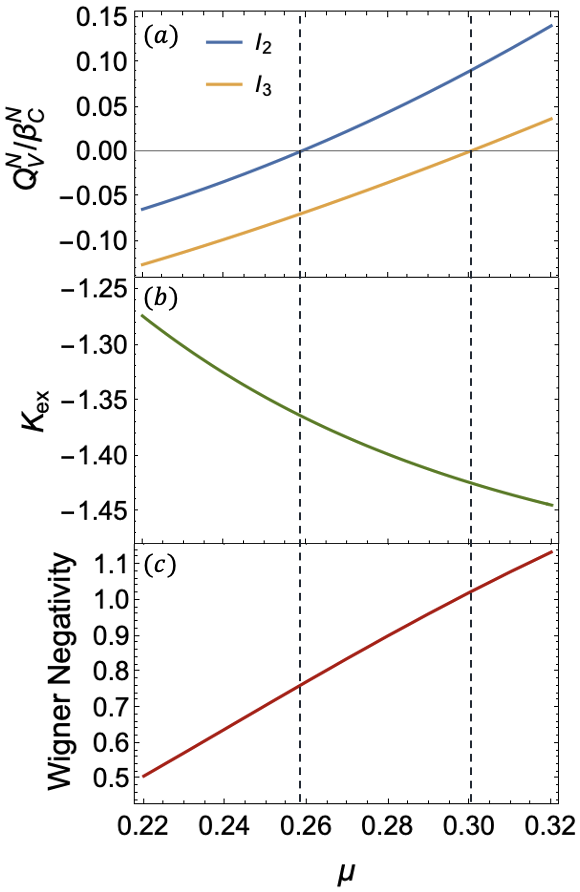}
	\end{center}
	\caption{Comparison between $I_2$ and $I_3$ violation and non-Gaussian properties of OAT states with $N=50$. As a function of the OAT interaction time $\mu$ we present: (a) relative quantum violation $Q_V^N/\beta_V^N$ of the 2nd-order PIBI $I_2$ (blue) and 3rd-order PIBI $I_3$ (orange), (b) excess kurtosis $K_{ex}$, and (c) Wigner negativity. Dashed vertical lines indicate the critical values of $\mu$ where the PIBIs violation vanishes.}
	\label{SI_nonGaussianity}
\end{figure}

Furthermore, for the highly non-Gaussian state maximising the quantum violation of $\hat{I}_4$ (whose Wigner function is presented in Fig.~\ref{fig:3} in the main text), we find an excess kurtosis of $K_{ex}=-1.94$ and a Wigner negativity of $\mathcal{N}=1.41$. Note that these numbers are much larger than the one for the OAT states detected by $I_2$ (see Fig.~\eqref{SI_nonGaussianity}). 

As we have discussed in the main text, Bell operators $\hat{I}_K(\phi_0,\theta_0,\phi_1,\theta_1)$ are functions of four angles for a given state. For $\hat{I}_4$ we fix the three angles $\phi_0,\theta_0,\phi_1$ to their optimal value while leaving $\theta=\theta_1$ to be variable, and show the relative quantum violation as a function of $\theta$ in Fig~\ref{SI_MeasureI2I4}. For the same sates and optimized measurement directions, we plot in the same figure the violation achievable by $I_2$. Clearly, we can see that the PIBI $I_4$ gives a relative quantum violation of $Q_V^N/\beta_C^N=-0.1390$ for a state where $I_2$ fails to reveal Bell correlations.

\begin{figure}[t]
	\begin{center}
	\includegraphics[width=80mm]{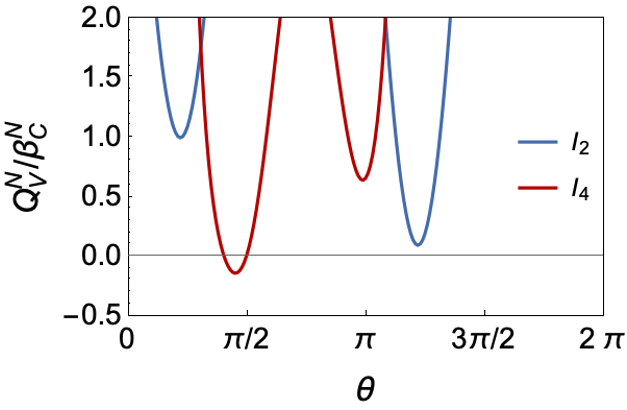}
	\end{center}
	\caption{Relative quantum violation of the 2nd-order PIBI $I_2$ (blue) and 4th-order PIBI $I_4$ (red) for $N=50$ eigenstate corresponding to minimum eigenvalue of $\hat{I}_4$ operator. For a given state, Bell operators $\hat{I}_K(\phi_0,\theta_0,\phi_1,\theta_1)$ are associated with four measurement angles. Here, we restricted three angles of them $\phi_0,\theta_0,\phi_1$ to be optimal ones corresponding to minimum relative quantum violation, and plot $Q_V^N/\beta_C^N$ changes with the measurement angle $\theta=\theta_1$.}
	\label{SI_MeasureI2I4}
\end{figure}

\end{widetext}

\end{document}